\documentclass[aps,prd,preprint,eqsecnum,tightenlines,floats,groupedaddress,showpacs,superscriptaddress, nofootinbib]{revtex4-1}

\pdfoutput=1  
 
\ifx\pdfoutput\undefined
\usepackage[dvips,bookmarks=false]{hyperref}	
\else
\usepackage{hyperref}	
\fi
\hypersetup{colorlinks,bookmarksopen,bookmarksnumbered,citecolor=blue,
linkcolor=black,pdfstartview=FitH,urlcolor=blue}

\usepackage{amssymb} 
\usepackage{hyperref} 
\usepackage{graphicx}
\usepackage{subfigure}
\usepackage{amsmath}
\usepackage{xcolor}
\usepackage{subfigure}
\usepackage[normalem]{ulem}

\usepackage{hyperref}	
\hypersetup{colorlinks,bookmarksopen,bookmarksnumbered,citecolor=blue, linkcolor=black,pdfstartview=FitH,urlcolor=blue}

\newcommand{\e}{\mathrm{e}}

\newcommand{\bb}{\begin{equation}}
\newcommand{\ee}{\end{equation}}
\newcommand{\bq}{\begin{eqnarray}}
\newcommand{\eq}{\end{eqnarray}}

\newcommand{\gag}{g_{a\gamma}}

\newcommand{\x}{\vec{x}}

\begin{document}


\date{\today}

\title{How to suppress exponential growth -- on the parametric resonance of photons 
in an axion background}

\author{Ariel Arza}
\affiliation{Institute for Theoretical and Mathematical Physics, Lomonosov Moscow State University (ITMP), 119991 Moscow, Russia}
\affiliation{Department of Physics, University of Florida, Gainesville, FL 32611, USA}

\author{Thomas Schwetz}
\affiliation{Institut f\"ur Kernphysik, Karlsruhe Institute of Technology (KIT),\\ 
Hermann-von-Helmholtz-Platz 1, 76344 Eggenstein-Leopoldshafen, Germany}

\author{Elisa Todarello}
\affiliation{Institut f\"ur Kernphysik, Karlsruhe Institute of Technology (KIT),\\
Hermann-von-Helmholtz-Platz 1, 76344 Eggenstein-Leopoldshafen, Germany}

\begin{abstract}

Axion--photon interactions can lead to an enhancement of the electromagnetic field by parametric resonance in the presence of a cold axion background, for modes with a frequency close to half the axion mass. In this paper, we study the role of the axion momentum dispersion as well as the effects of a background gravitational potential, which can detune the resonance due to gravitational redshift. We show, by analytical as well as numerical calculations, that the resonance leads to an exponential growth of the photon field only if (a) the axion momentum spread is smaller than the inverse resonance length, and (b) the gravitational detuning distance is longer than the resonance length. For realistic parameter values, both effects strongly suppress the resonance and prevent the exponential growth of the photon field. In particular, the redshift due to the gravitational potential of our galaxy prevents the resonance from developing for photons in the observable frequency range, even assuming that all the dark matter consists of a perfectly cold axion condensate. For axion clumps with masses below $\sim 10^{-13}\, M_\odot$, the momentum spread condition is more restrictive, whereas, for more massive clumps, the redshift condition dominates.

\end{abstract}

\maketitle

\tableofcontents

\section{Introduction}

The QCD axion~\cite{Peccei:1977ur,Peccei:1977hh,Weinberg:1977ma,Wilczek:1977pj}
is one of the leading candidates to compose the dark matter in the universe. Its production in the early universe by non-thermal mechanisms, such as the vacuum re-alignment mechanism and the decay of topological defects, makes it abundant and cold enough to fit the dark matter requirements \cite{Preskill:1982cy, Abbott:1982af, Turner:1983he, Dine:1982ah, Sikivie:2006ni, Marsh:2015xka}. The QCD axion is characterized by a single parameter, $f_a$, that corresponds to the energy scale at which the Peccei-Quinn symmetry is broken. Thus, its mass, which is generated by non-perturbative effects, and all its couplings to Standard Model (SM) particles scale like $\sim1/f_a$. On the other hand, several extensions to the SM, and especially string theory, predict the existence of pseudo-scalar fields with similar features as those of the QCD axion \cite{Arvanitaki:2009fg, Ringwald:2012hr}, so-called axion-like particles (ALPs). They have the same type of interactions with the SM, and, most importantly, they are also dark matter candidates \cite{Arvanitaki:2009fg, Ringwald:2012hr, Arias:2012az}. The difference between QCD axions and ALPs lies in the fact that, for the latter, there is no direct relationship between the symmetry breaking scale and the mass. From now on, we will use the word axion to refer to the QCD axion or ALPs.

The coupling of the axion to two photons  is very interesting from the phenomenological point of view~\cite{Sikivie:1983ip}. For the photon-axion interaction, the Lagrangian density is given by
\bb
{\cal L}_{a\gamma\gamma}=-\frac{1}{4}F_{\mu\nu}F^{\mu\nu}+\frac{1}{2} \partial_\mu a\partial^\mu a - \frac{1}{2} m_aa^2+\frac{1}{4}\gag aF_{\mu\nu}\tilde F^{\mu\nu} \label{iL1}  \enspace,
\ee
where $a$ is the axion field, $m_a$ the axion mass, ${F_{\mu\nu}=\partial_\mu A_\nu-\partial_\nu A_\mu}$ the electromagnetic field strength tensor, with
$A_\mu$ being the electromagnetic vector field,
$\tilde F^{\mu\nu}=\frac{1}{4}\epsilon^{\mu\nu\alpha\beta}F_{\alpha\beta}$, and $\gag$ is the coupling constant with mass dimension $(-1)$. Based on this coupling, several experiments and ideas have been developed to search for the axion, see \cite{Irastorza:2018dyq,Sikivie:2020zpn} for reviews. Most of them exploit the axion to photon transition when a static magnetic field is present. Even though none of these experiments has found an axion signal so far, the parameter space that remains to be explored is still huge. For the numerical estimates below, we will adopt a reference value for the axion--photon coupling constant of $\gag \simeq 10^{-11}\, {\rm GeV}^{-1}$, roughly a factor 6 below the current upper limit from the CAST experiment~\cite{Anastassopoulos:2017ftl}, valid for a wide range of axion masses.

Another consequence of the interaction in Eq.~\eqref{iL1} is the
possibility of axion decay into two photons, which can be resonantly
enhanced due to stimulated decay. This effect has been discussed early
on in the axion literature, e.g.~\cite{Preskill:1982cy, Abbott:1982af,
  Dine:1982ah, Tkachev:1986tr, Tkachev:1987cd}. More recently, many
authors have considered various aspects of spontanteous or stimulated
axion decay \cite{Kephart:1994uy,Masso:1997ru,Yoshida:2017ehj, Arza:2018dcy,
  Caputo:2018ljp,Caputo:2018vmy, McDonald:2019wou, Arza:2019nta,
  Sigl:2019pmj, Wang:2020zur, Chen:2020ufn}, including a cosmological context
\cite{Lee:1999ae,Mirizzi:2009nq,Alonso-Alvarez:2019ssa}, axion stars
\cite{Tkachev:2014dpa, Hertzberg:2018zte}, or superradiant axion
clouds around black holes \cite{Rosa:2017ury,Ikeda:2019fvj}.

In this work, we consider the propagation of photons in the presence of
a background axion field, focusing on the parametric resonance
enhancement of the photon field, happening if the photon frequency is close to
half the axion mass. This effect can be interpreted as stimulated
axion decay and can lead to an exponential growth of the photon field,
with potentially observable consequences, e.g.~\cite{Tkachev:1987cd,
  Arza:2018dcy, Tkachev:2014dpa, Hertzberg:2018zte,
  Sigl:2019pmj}.\footnote{An analogous effect for gravitational waves
  propagating through the axion background has been studied recently
  in \cite{Jung:2020aem, Sun:2020gem}.}  We study in detail the impact of a
non-trivial momentum distribution of the axion field (including
coherent axion clumps as well as random axion fields with a finite
coherence length) and the effect of gravitational redshift due to a
gravitational potential. Both effects can dramatically reduce the
development of the instability. The main purpose of the present work
is to establish conditions which need to be satisfied in order for the
resonance to be effective. 

Throughout this work, we ignore the
back reaction of the resonance on the axion field and treat the axion
as an external background field. This is justified as long as we are interested
in the question of whether an instability develops; ultimately, the
exponential growth will be cut off by backreactions on the axion
field. Recent considerations of this effect can be found in
\cite{Sawyer:2018ehf, Carenza:2019vzg}.  In order to isolate the
effects of the axion coherence length and gravitational redshift, we also
neglect possible effects of the effective photon mass in the plasma
\cite{Mirizzi:2009nq, Wang:2020zur} and the presence of external magnetic fields \cite{Masaki:2019ggg}.

The effect of axion momentum dispersion on the parametric resonance has been discussed by several authors in the past, for instance \cite{Tkachev:2014dpa, Hertzberg:2018zte, Arza:2018dcy, Wang:2020zur}. In agreement with previous authors, we find that the resonance is suppressed if the momentum spread is large compared to the inverse resonance length. We study this phenomenon with numerical and analytical methods, assuming either a coherent clump of axions
(such as, for instance, axion stars), where the momentum spread is set by the inverse of the size of the object, or a random axion field with a given coherence length. The impact of gravitational redshift has, to our knowledge, not been discussed quantitatively in the literature. Some estimates have been given recently in Ref.~\cite{Wang:2020zur}. We show that, under the hypothesis of a coherent, monochromatic axion background field, the gravitational redshift can shift photon modes out of the resonance band. The resonance can only develop if the detuning distance is much larger than the resonance distance. Since the width of the resonance band is extremely tiny for typical parameter values, very weak gravitational potentials suffice to destroy the resonance. For instance, in Ref.~\cite{Sigl:2019pmj}, strong sensitivities to the axion--photon coupling have been derived under the assumption that photons entering our galaxy experience a parametric resonance. However, we will show below that the redshift effect due to the gravitational potential of our galaxy prevents the exponential growth of the photon field. 

The outline of the remainder of the paper is as follows. In Section~\ref{sec:essentials}, we review the axion--photon parametric resonance, assuming that the axion is a plane wave with fixed momentum. In Section~\ref{sec:dispersion}, we generalize to the case of a superposition of momentum modes for the axion background. This problem is treated numerically in Section~\ref{sec:num}, where we introduce our simulation framework. In Section~\ref{sec:redshift}, we discuss the impact of a gravitational potential, and we summarize and discuss our results in Section~\ref{sec:disc}. Supplementary material and technical details are given in Appendices~\ref{eikonal}, \ref{wkb}, \ref{num_detail}, \ref{steady_detail}.

\section{Essentials of the photon-axion parametric resonance}\label{sec:essentials}

In this Section, we review the parametric resonance caused by the axion--photon interactions.
To set the stage, we take the axion field to be a monochromatic wave with a non-zero momentum. This generalizes previous discussions of the homogeneous case with zero-momentum axions, see e.g., Refs.~\cite{Arza:2018dcy,Hertzberg:2018zte}. A similar discussion of parametric resonance is presented in~\cite{Arza:2019kab}, in the context of a scalar field with repulsive self-interactions.

First, we write the equations of motion derived from the Lagrangian density (\ref{iL1}):
\bq
\partial_\mu F^{\mu\nu} &=&  \gag \partial_\mu a\tilde F^{\mu\nu} \label{eqcov1}
\\
\left(\partial_\mu\partial^\mu+m_a^2\right)a &=& \frac{1}{4} \gag F_{\mu\nu}\tilde F^{\mu\nu} \enspace. \label{eqcov2}
\eq
In the Coulomb gauge ($\vec\nabla\cdot\vec A=0$), these equations can be written as %
\bq
\nabla^2A_0 &=& - \gag \vec\nabla a\cdot\nabla\times\vec A \label{eq01}
\\
\left(\partial_t^2-\nabla^2\right)\vec A+\partial_t\vec\nabla A_0 &=&  \gag \vec\nabla a\times(\partial_t\vec A+\vec\nabla A_0)- \gag \partial_ta\vec\nabla\times\vec A \label{eq1}
\\
\left(\partial_t^2-\nabla^2+m_a^2\right)a &=&  \gag (\partial_t\vec A+\vec\nabla A_0)\cdot\vec\nabla\times\vec A \enspace. \label{eq1ax}
\eq
As mentioned above, we assume for the moment a monochromatic axion field of the form
\begin{equation}
a(t, \vec x)={\sqrt{2\rho_a}\over m_a}\sin\left(\omega_at-\vec p\cdot\vec x\right) \enspace. \label{axionfield1}
\end{equation}
Here, $\rho_a$ is the axion energy density, $\vec p$ the axion wave vector and $\omega_a=\sqrt{p^2+m_a^2}$ the axion angular frequency. Neglecting the backreaction on the axion, we can consider $\rho_a$ as constant, and ignore Eq.~(\ref{eq1ax}). We  also assume non-relativistic axions, i.e.~$\omega_a\approx m_a \gg p$, and hence  neglect spatial derivatives of $a$. However, we keep $\vec p$ in the phase of the axion field. Thus, Eq.~(\ref{eq01}) has no source, and therefore we can set $A_0=0$. All these valid assumptions reduce the problem to solve to
\begin{equation}
\left(\partial_t^2-\nabla^2\right)\vec A=- \gag \sqrt{2\rho_a}\cos\left(\omega_at-\vec p\cdot\vec x\right)\vec\nabla\times\vec A \enspace. \label{eq2}
\end{equation}

We write $\vec A$ as a Fourier expansion 
\bb
\vec A(t, \vec x)=\sum_{\lambda=1}^2\int{d^3k\over(2\pi)^3}\,e^{i\vec k\cdot\vec x}\,\hat\epsilon_{\vec k,\lambda}A_{\vec k,\lambda}(t) +\text{c.c.} \enspace, \label{eAexp1}
\ee
where $\hat\epsilon_{\vec k,\lambda}$ are the linear polarization vectors satisfying $\hat\epsilon_{\vec k,\lambda}\cdot\hat\epsilon_{\vec k,\lambda'}=\delta_{\lambda,\lambda'}$ and $\hat\epsilon_{\vec k,1}\times\hat\epsilon_{\vec k,2}=\hat k$. We plug (\ref{eAexp1}) into Eq.~(\ref{eq2}), then multiply by $e^{-i\vec k'\cdot\vec x}$, integrate over space and project onto $\hat\epsilon_{\vec k',\lambda'}$. Moreover, we keep only terms that are relevant for the stimulated axion decay, which are the terms that can resonate.
By momentum conservation during the axion decay $\vec p \to \vec k + (\vec p - \vec k)$, the modes involved in the process are $A_{\vec k, \lambda}$ and $A_{\vec p-\vec k,\lambda}^*$.
For each mode $\vec k$ and polarization $\lambda$, we get
\begin{align}
(\partial_t^2+k^2)A_{\vec k,\lambda} = i \gag \sqrt{\rho_a\over2}\,\hat\epsilon_{\vec k,\lambda}\cdot\sum_{\lambda'=1}^2 & (\vec p-\vec k)\times\hat\epsilon_{\vec p-\vec k,\lambda'}^*A_{\vec p-\vec k,\lambda'}^*e^{-i\omega_at} \enspace. \label{eq3}
\end{align}
Eq.~(\ref{eq2}) was derived under the assumption that the contribution of the spatial derivatives of the axion field, of order the axion momentum $\vec p$, is small.
Hence, we neglect sub-leading contributions of order $\vec p$, considering also that photon momenta is of order of the resonance frequency $k\simeq m_a/2 \gg p$. We define $\vec q=\vec p-\vec k$ and approximate $q\simeq k(1+{\cal O}(p/k))$ and $\hat\epsilon_{\vec q,\lambda}^*\simeq\hat\epsilon_{-\vec k,\lambda}^*(1+{\cal O}(p/k))$.
Eq.~(\ref{eq3}) becomes the coupled system
\bq
(\partial_t^2+k^2)A_{\vec k,1} &=& -i \gag \sqrt{\rho_a\over2}\,q\,\hat\epsilon_{\vec k,1}\cdot\hat\epsilon_{\vec q,1}^*e^{-i\omega_at}A_{\vec q,2}^* \label{eqk1}
\\
(\partial_t^2+q^2)A_{\vec q,2}^* &=& -i \gag \sqrt{\rho_a\over2}\,k\,\hat\epsilon_{\vec k,2}\cdot\hat\epsilon_{\vec q,2}^*e^{i\omega_at}A_{\vec k,1}  \enspace. \label{eqq1}
\eq

To solve Eqs.~(\ref{eqk1}) and (\ref{eqq1}), we look for solutions where the fields $A_{\vec k,1}$ and $A_{\vec q,2}^*$ oscillate close to the frequencies $k$ and $q$, respectively. In the stimulated axion decay, the two polarizations propagate in almost opposite directions, since $|\vec p\,| \ll |\vec k|$. We can then make the ansatz ${A_{\vec k,1}(t)=\alpha_{\vec k}(t)e^{-ikt}e^{i\epsilon_{\vec k}t/2}}$ and $A_{\vec q,2}^*(t)=\beta_{\vec q}(t)e^{iqt}e^{-i\epsilon_{\vec k}t/2}$, where $|\dot\alpha_{\vec k}|\ll k|\alpha_{\vec k}|$, $|\dot\beta_{\vec q}|\ll q|\beta_{\vec q}|$ and 
\bb
\epsilon_{\vec k}=k+ q - \omega_a  =
2k-m_a - p \cos\left(\varphi_{\vec k}\right) + \mathcal{O}(p^2)  \enspace,
\label{eeps1}
\ee
with $\varphi_{\vec k}$ being the angle between $\vec k$ and $\vec p$.
We introduced the factors $e^{\pm i\epsilon_{\vec k}t/2}$ as a matter of convenience. Thus, Eqs. (\ref{eqk1}) and (\ref{eqq1}) transform to
\begin{eqnarray}
\left(\partial_t+i\epsilon_{\vec k}/2\right)\alpha_{\vec k}&=&-\sigma\beta_{\vec q} \label{eeak1}
\\
\left(\partial_t-i\epsilon_{\vec k}/2\right)\beta_{\vec q}&=&-\sigma\alpha_{\vec k} \enspace, \label{eebk1}
\end{eqnarray}
where we defined 
\bb
\sigma={ \gag \over2}\sqrt{\rho_a\over2} \enspace. \label{esigma1}
\ee
Applying the operator $\partial_t-i\epsilon_{\vec k}/2$ to (\ref{eqk1}) and $\partial_t+i\epsilon_{\vec k}/2$ to (\ref{eebk1}), we get
\begin{equation}
\left(\partial_t^2-s_{\vec k}^2\right)\left(
\begin{array}{cc}
\alpha_{\vec k}
\\
\beta_{\vec q}
\end{array}
\right)=0  \enspace,\label{eqkbk}
\end{equation} 
where
\begin{equation}
s_{\vec k}=\sqrt{\sigma^2-\epsilon_{\vec k}^2/4}  \enspace. \label{esk1}
\end{equation}
It is clear that for $s_{\vec k}^2>0$ there is an instability that causes the amplitudes grow exponentially in time. It is the well known phenomenon of parametric resonance. The resonance affects the electromagnetic field for modes that satisfy
\bb
-2\sigma<\epsilon_{\vec k}<2\sigma  \enspace,\label{band}
\ee
or, using Eq.~\eqref{eeps1},
\bb
\frac{m_a}{2}(1+v_a\cos\varphi_{\vec k}) - \sigma < k <
\frac{m_a}{2}(1+v_a\cos\varphi_{\vec k}) + \sigma  \enspace,
\ee
with $v_a = p/m_a \ll 1$ being the axion velocity. Hence, the width of the resonance band is equal to $2\sigma$. The complete solution of the coupled system Eqs.~(\ref{eeak1}, \ref{eebk1}) is
\begin{eqnarray}
\alpha_{\vec k}(t) &=& \alpha_{\vec k}(0)\cosh(s_{\vec k}t)-{1\over s_{\vec k}}\left(\sigma\beta_{\vec q}(0)+{i\over2}\epsilon_{\vec k}\alpha_{\vec k}(0)\right)\sinh(s_{\vec k}t) \label{solak1}
\\
\beta_{\vec q}(t) &=& \beta_{\vec q}(0)\cosh(s_{\vec k}t)-{1\over s_{\vec k}}\left(\sigma \alpha_{\vec k}(0)-{i\over2}\epsilon_{\vec k}\beta_{\vec q}(0)\right)\sinh(s_{\vec k}t)  \enspace. \label{solbk1}
\end{eqnarray}
At late times, this leads to exponential growth $\propto \exp(s_{\vec k}t)$, with the maximal growth rate at the center of the resonance band where $\epsilon_{\vec k} = 0$ and $s_{\vec k} = \sigma$. The quantity $\sigma$ sets the scale of the resonance, we denote it in the following as ``growth factor''. The inverse $\sigma^{-1}$ corresponds to the resonance time or equivalently resonance length. From Eq.~\eqref{eeps1}, we obtain for the photon frequency at the center of the resonance 
\bb
k_{\rm res} = \frac{m_a}{2} (1 + v_a\cos\varphi_{\vec k})  \enspace. 
\ee
We see that a finite axion velocity shifts the center of the resonance band from $m_a/2$. This will be crucial for the gravitational redshift effect discussed in Section~\ref{sec:redshift}. Since, photon waves traveling in opposite directions are produced, for a given photon momentum, we expect two resonance peaks located at $m_a/2(1 \pm v_a |\cos\varphi_{\vec k}|)$.

Let us give some typical numerical estimates in order to get a feeling for the scales involved. Assuming axion-photon couplings not too far below current limits, and taking as a benchmark for the axion density the local dark matter density in our galaxy, we find from Eq.~\eqref{esigma1} 
\bb
\sigma \simeq 6\times 10^{-24}\,{\rm eV}
\left(\frac{\gag}{10^{-11} \,{\rm GeV}^{-1}}\right)
\left(\frac{\rho_a}{0.4 \,{\rm GeV/cm}^{3}}\right)^{1/2} \label{sigma_value} \enspace.
\ee
Converting this reference scale, $\sigma_{\rm ref} \equiv 
6\times 10^{-24}\,{\rm eV}$,
into a distance we obtain, for those fiducial numbers,
$1/\sigma_{\rm ref} \simeq 1$~pc. For an axion mass of $m_a = 10^{-5}$~eV, the relative width of the resonance is $2\sigma_{\rm ref}/m_a \simeq 10^{-18}$. As we will see in the following, it is this extremely tiny width of the resonance band that makes it difficult for the resonance to develop in realistic situations.

\section{Momentum dispersion effects} \label{sec:dispersion}

Let us now extend our analysis to the scenario where the axion field is composed by a superposition of momentum modes. We write the (real) axion field as a superposition of plane waves:
\begin{equation}\label{eq:a_discrete}
  a(t, \vec x) = \frac{1}{m_a} \sqrt{\frac{\langle\rho_a\rangle}{2}}e^{-im_at}
  \sum_p b_{\vec{p}}\, e^{i\vec{p}\cdot\vec{x}} + {\rm c.c.}  \enspace.
\end{equation}
Here, $b_{\vec{p}}$ are dimensionless coefficients normalized as
\begin{equation}\label{eq:norm}
  \sum_{\vec{p}} | b_{\vec{p}}\, |^2 = 1  \enspace,
\end{equation}
and $\langle\rho_a\rangle$ is the axion energy density averaged over a time $T \gg 1/m_a$ and over a large volume $V = L^3$. The local energy density is given by 
\begin{equation}
  \rho_a(t, \vec x) \approx \frac{1}{2} \left[ (\partial_t a)^2 + m_a^2 a^2\right]
  \approx m_a^2 a^2(t, \vec x)  \enspace,
\end{equation}
where we have neglected the momentum contribution to the energy density. The finite size of the volume implies a finite momentum resolution, $\Delta = 2\pi/L$, which leads to the discrete sum over the momentum modes in Eq.~\eqref{eq:a_discrete}.

\begin{figure}[t]
  \centering
  \includegraphics[width=0.75\textwidth]{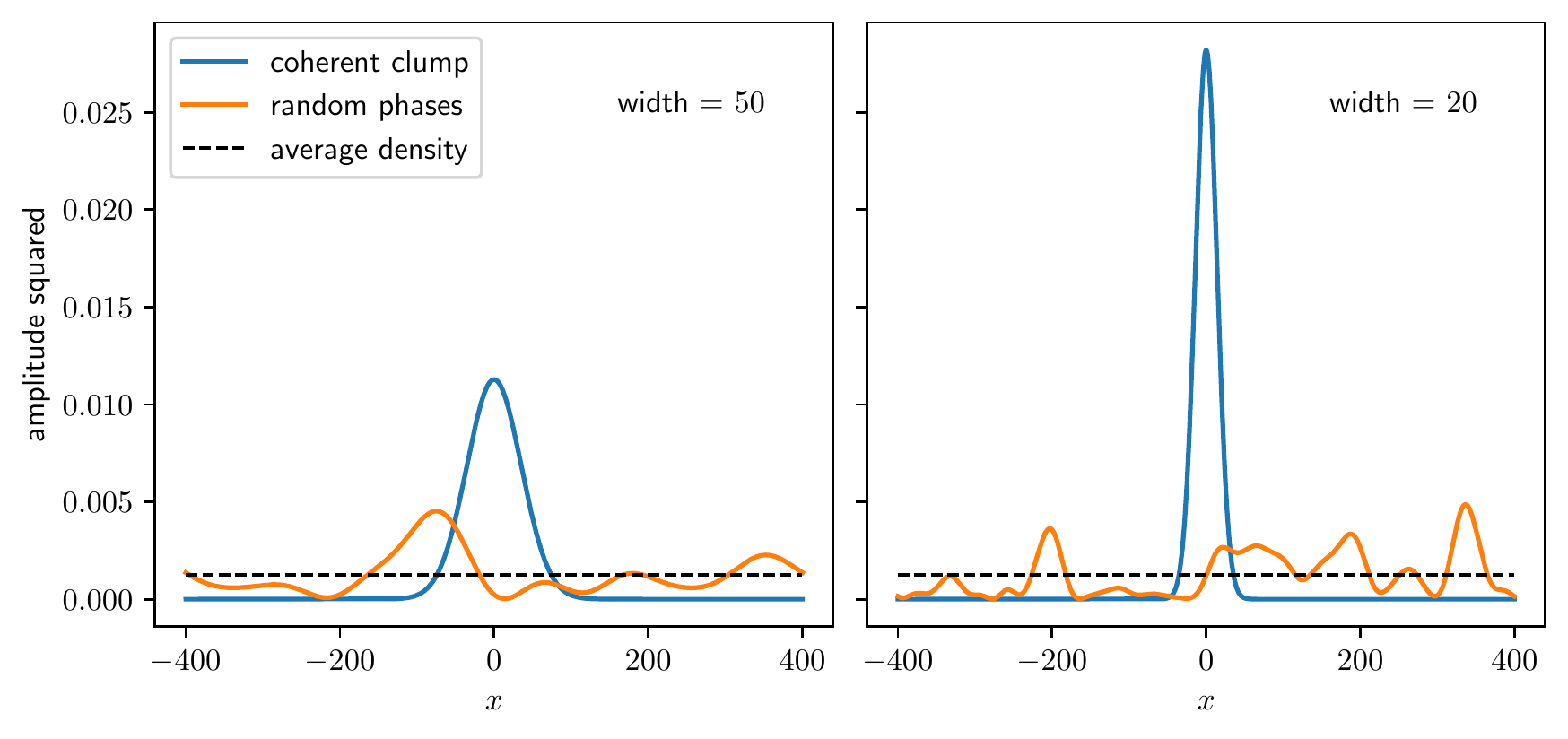}
  \caption{Time averaged energy density for Gaussian momentum
    distributions with $\delta_p = 1/50$ (left) and 1/20 (right), see Eq.~\eqref{eq:rho_m_gauss}. 
The space coordinate $x$ has arbitrary units and $\delta_p$ has units of inverse length.    
For the blue curves, we take $\varphi_p= 0$. For the orange curves, we assign random values to $\varphi_p$ for each $p$. Distributions are normalized to 1. The average density is shown as a dashed line.}
    \label{fig:dens}
\end{figure}

As a generic example, we consider in the following a 
Gaussian wave packet with width $\delta_p$ in
momentum space. Our normalization implies
\begin{equation}\label{eq:wp}
|b_p|^2 = \left(\frac{\Delta}{\sqrt{\pi}\delta_p}\right)^3 e^{-\frac{p^2}{\delta_p^2}}  \enspace.
\end{equation}
This expression holds for $\Delta \ll
\delta_p$. If $\Delta \gtrsim \delta_p$, we replace it by integrating the Gaussian over the bin width and imposing the normalization condition in Eq.~\eqref{eq:norm}. The time averaged energy density at a given point $\vec x$ becomes
\begin{equation}\label{eq:rho_m_gauss}
  \overline{\rho}_a(\vec{x}) =
  \langle\rho_a\rangle 
  \left|\sum _p b_p \, e^{i\vec{p}\cdot\vec{x}}  \right|^2 =
  \langle\rho_a\rangle \left(\frac{\Delta}{\sqrt{\pi}\delta_p}\right)^3
  \left|\sum _p e^{-\frac{p^2}{2\delta_p} + i\varphi_p} \, e^{i\vec{p}\cdot\vec{x}}  \right|^2 \enspace.
\end{equation}
In the second equality, we have defined the phases of $b_p$ as
$b_p = |b_p| e^{i\varphi_p}$. 
For $\varphi_p=0$, this corresponds to a localized wave packet, whereas
for random phases, we obtain a fluctuating energy density with
coherence length $\delta_p$. This is illustrated, for one dimension, in Fig.~\ref{fig:dens}.

We Fourier expand the photon vector potential as in Eq.~(\ref{eAexp1}), but now we use helicity eigenstates $\hat{e}_{\vec q, \pm}$ that satisfy $i\vec{q} \times
\hat{e}_{\vec q, \pm} = \pm |\vec{q}\,| \hat{e}_{\vec q, \pm}$. In this case, the equations for the two helicities decouple. We suppress
the helicity index for ease of notation in the following.
With the axion field from Eq.~\eqref{eq:a_discrete}, 
the differential equation for the time dependent Fourier coefficients, analogous to Eq.~\eqref{eq3}, becomes 
\begin{align}\label{eq:mathieu_disc}
  (\partial_t^2 + k_r^2)A_r(t) = \pm i\tilde g m_a
\sum_s
  \left[b_s e^{im_at}|k_r-p_s|A_{r-s}(t) - 
       b_s^* e^{-im_at}|k_r+p_s|A_{r+s}(t) \right] \enspace,
\end{align}
where we have restricted to one space dimension, discretized
momentum as described above, and defined the dimensionless parameter
\begin{align}
  \tilde g \equiv \frac{\gag}{m_a}
  \sqrt{\frac{\langle\rho_a\rangle}{2}} \ll 1 \,.
\end{align}
The sign $+$ ($-$) on the RHS of (\ref{eq:mathieu_disc}) corresponds to the polarization $\hat e_{\vec q,+}$ ($\hat e_{\vec q,-}$). From now on, we will only take into account the polarization $\hat e_{\vec q,+}$. The final results will be identical for both polarizations.
We consider $N$ (with $N$ odd) momentum modes centered around the
resonance momentum $m_a/2$ and we use the same grid for the axion momentum:
\begin{equation}\label{eq:indices}
  k_r = \frac{m_a}{2} + r \Delta \,,\quad r = -\frac{N-1}{2},...,\frac{N-1}{2} \,, \quad p_s = s \Delta  \enspace.    
\end{equation}
We adopt the obvious notation $A_r(t) \equiv A(t,k_r)$.  We
address Eq.\eqref{eq:mathieu_disc} by two different methods, which we
briefly describe in the following.

\bigskip

{\bf Method A: WKB approximation and frequency matching.} 
Using Eq.~\eqref{eq:indices} and after a re-labeling of the index $s$ on the
r.h.s., Eq.~\eqref{eq:mathieu_disc} can be written as
\begin{align}\label{eq:disc2}
  \left[\partial_t^2 + \left(\frac{m_a}{2}+ r\Delta\right)^2\right] A_r(t) = i\tilde g m_a
\sum_s \left(\frac{m_a}{2}+ s\Delta\right)
  \left[b_{r-s} e^{im_at}  - b_{-r+s}^* e^{-im_at}\right] A_{s}(t)  \enspace.
\end{align}
Using the ansatz $A_r = \alpha_r(t) e^{ik_rt} + \text{c.c.}$, with
$\alpha_r(t)$ slowly varying, and keeping only the terms which
resonate (``frequency matching''), we find to first order
\begin{equation}
  \left(\frac{m_a}{2}+ r\Delta\right) e^{i r \Delta t} \partial_t\alpha_r =
  \frac{\tilde g m_a}{2} \sum_s 
  \left(\frac{m_a}{2}+ s\Delta\right) b_{r-s} e^{-i s \Delta t} \alpha_s^*  \enspace.
\end{equation}
This system can be further simplified by removing the time dependent
exponentials with the change of variables $\beta_r(t) = (m_a/2 +
r\Delta) e^{i r \Delta t} \alpha_r(t)$. Under the condition $\Delta
\ll m_a$, $\beta_r(t)$ is slowly varying in the same
sense as $\alpha_r(t)$. The equations become
\begin{equation}\label{eq:beta1}
  \partial_t\beta_r = ir\Delta \beta_r + \frac{\tilde g m_a}{2}\sum_s b_{r-s} \beta_s^*  \enspace.
\end{equation}
By considering the complex conjugate of these equations, we obtain a
closed system of 1st order differential equations:
\begin{equation}\label{eq:beta2}
  \partial_t \left(\begin{array}{c} \beta \\ \beta^* \end{array}\right) =
  {\cal M} \left(\begin{array}{c} \beta \\ \beta^* \end{array}\right)  \enspace,
\end{equation}
with ${\cal M}$ being a time-independent square matrix of order $2N$ whose
elements follow from Eq.~\eqref{eq:beta1}.  This system is solved by
linear combinations of functions of the form $e^{\lambda_r t}$, with
$\lambda_r$ being the eigenvalues of the matrix ${\cal M}$. An eigenvalue
with a positive real part will lead to an exponentially growing
solution, signaling the instability. We calculate the eigenvalues of
${\cal M}$ numerically. The eigenvalue with the largest real part will
dominate at late times.

\bigskip

{\bf Method B: Floquet analysis.} 
A standard method for the solution differential equations with periodic coefficients 
is the Floquet analysis, see
e.g.~\cite{mathieu, cesari} for general reviews and 
\cite{Hertzberg:2018zte} for an application in the axion context. Eq.~\eqref{eq:mathieu_disc} can be converted into a 1st order differential equation $\partial_t x = {\cal M}' x$, with $x =
(B,A)^T$, $B_r = \partial_t A_r$, where $x(t)$ and ${\cal M}'(t)$ are a column
vector and square matrix of order $2N$, respectively, where the
elements of ${\cal M}'$ follow from Eq.~\eqref{eq:mathieu_disc}. Using that
${\cal M}'(t)$ is periodic, ${\cal M}'(t) = {\cal M}'(t + 2\pi/m_a)$, we can obtain the
long-term behavior by solving the system for one period, for a
complete set of initial conditions, obtaining an evolution matrix $C$
which evolves the system over one period. Let us denote by $\lambda'_r$
the (complex) eigenvalues of $C$.  Then, the state of the system after many
periods at time $t=n 2\pi/m_a$ is determined by ${\lambda'_r}^n = e^{\sigma_r
  t}$, with $\sigma_r = m_a/(2\pi) \log(\lambda'_r)$ being called Floquet
exponent. Hence, eigenvalues with absolute value $>1$ will lead to an
unstable mode that grows exponentially.

\bigskip

\begin{figure}[t!]
  \centering
  \includegraphics[width=\textwidth]{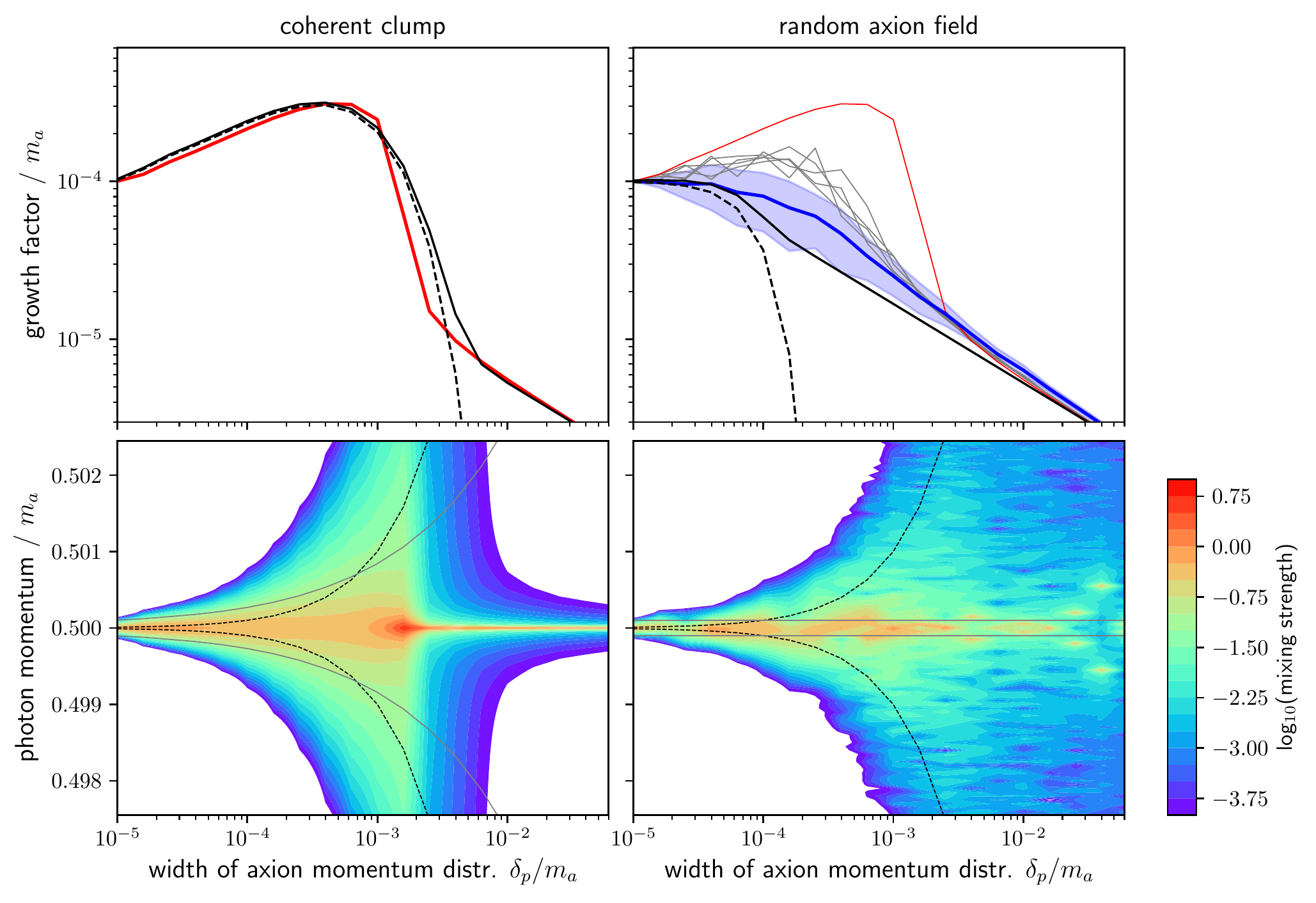}
  \caption{Maximal growth factor (upper panels) and mixing of photon
    momenta with the eigenvector corresponding to the maximal growth
    (lower panels) as a function of the width of a Gaussian axion momentum distribution. The color code indicates log$_{10}$ of the mixing strength. The left panels correspond to the case of a coherent axion clump, whereas the right panels assume random phases for each axion momentum mode. We take $\tilde g = 2\times 10^{-4}$ and
    $N=99$ momentum bins with width $\Delta/m_a = 5\times
    10^{-5}$. In the lower panels, the gray curves indicate the resonance band
    (Eq.~\eqref{eq:mu-bar}, see text), and the dashed curves
    correspond to $k=m_a/2\pm \delta_p$. The
    right upper panel shows the mean and the one standard deviation
    band for the growth factor obtained from method A (blue curve and
    shaded region), thin grey curves correspond to five random
    realizations from method B; the thin red curve corresponds to the
    coherent clump (same as left upper panel). The lower-right panel
    shows the mixing for one arbitrary example of the five random
    realizations. The solid and dashed black curves in the upper
    panels show the growth factors estimated according to
    Eq.~\eqref{eq:mu_est}.
    \label{fig:floquet}}
\end{figure}

Both methods give consistent results, i.e., the largest real part of
the eigenvalues from Method A agrees with the largest Floquet exponent
from Method B. The agreement is perfect for the coherent clump, while
some deviations appear for random axion fields, see
Fig.~\ref{fig:floquet}.

In Fig.~\ref{fig:floquet}, we show the results of such an analysis. We
assume a Gaussian momentum distribution and show the results as a
function of the width of the Gaussian. The left panels correspond to a
coherent axion clump with real $b_r$, while the right panels
assume random phases, $b_r = |b_r|e^{i\varphi_r}$, with
$\varphi_r$ taking a random value for each mode $r$. The upper panels
show the growth factor corresponding to the largest Floquet exponent,
while the lower panels show the mixing strength of the photon momentum
modes with the eigenvector corresponding to the largest Floquet
exponent.  In the numerical analysis, we divide
Eq.~\eqref{eq:mathieu_disc} by $m_a^2$ and consider all dimensionful
quantities in units of the axion mass. For Fig.~\ref{fig:floquet}, we
have used $N=99$, $\Delta/m_a = 5\times 10^{-5}$, and $\tilde g = 2\times
10^{-4}$.  

Let us discuss the results shown in the figure.
For very small width, $\delta_p \lesssim 10^{-5}$, the axion
distribution is effectively homogeneous, with all $b_r = 0$
except $|b_0| = 1$. In this situation, the coherent and random cases
become identical, and we recover the well-known results for the
homogeneous axion with a growth factor equal to $\tilde g m_a/2$, and the
resonance frequency at $m_a/2$.

For increasing $\delta_p$, more and more axion momentum bins become
populated. We can understand the behavior in the following way.  The
eigenvalues of the matrix $\mathcal{M}$ in Eq.~\eqref{eq:beta2} are
determined from the matrix with elements
\begin{equation}\label{eq:beta3}
\frac{\tilde g m_a}{2} b_{r-s}   \enspace,
\end{equation}
appearing in
Eq.~\eqref{eq:beta1}. As more and more components of $b_{r-s}$
contribute, we can obtain an order of magnitude estimate for the
largest eigenvalue as
\begin{equation}\label{eq:mu-bar} 
  \sigma \equiv \frac{\tilde g m_a}{2} \left|\sum_r b_r\right|
  = \frac{\gag}{2} \sqrt{\frac{\overline\rho_a (0)}{2}}  \enspace.
\end{equation}
In the second equality, we have used the definition of $\tilde g$ and
the time-averaged axion density from Eq.~\eqref{eq:rho_m_gauss}. We obtain a generalization of the growth factor for the monochromatic axion case, Eq.~\eqref{esigma1}. In
the case of the coherent clump, it corresponds to the local density at
the peak of the clump, whereas, for the random field, we have
$\overline\rho_a (\vec{x}) \approx \langle \rho_a \rangle$ on average.  In
analogy to the monochromatic axion case, we can take $m_a/2 \pm
\sigma$ as the effective width of the resonance band. This is
illustrated in the lower panels of Fig.~\ref{fig:floquet} by the grey
curves. We see that, in the coherent case, the momentum modes within
the resonance band add up coherently, and we notice an increased growth
factor. While in the random phase case, the growth factor remains
similar to the homogeneous case, up to random fluctuations of order a factor of 2.  This is consistent with the interpretation in
Eq.~\eqref{eq:mu-bar} in terms of the axion density
$\overline\rho(0)$, which increases for a Gaussian clump at constant
mass when increasing $\delta_p$, while it remains constant for the
random field, c.f.~Fig.~\ref{fig:dens}.

If $\delta_p$ becomes larger than the resonance band, all coefficients
$|b_{r-s}|$ within the band become equal (the exponentials in
Eq.~\eqref{eq:wp} are $\approx 1$). Hence, the matrix in
Eq.~\eqref{eq:beta3} becomes singular; there is only one non-zero
eigenvalue given by
\begin{equation}
\sigma_{\rm sing} = \frac{\tilde g m_a}{2} |b_0|  
= {\gag\over2} \sqrt{\frac{\langle \rho_a\rangle}{2}}
\left(\frac{\Delta}{\sqrt{\pi} \delta_p}\right)^{n/2}
={\gag\over2} \sqrt{\frac{M}{2}}
\frac{\Delta^n}{(2\pi\sqrt{\pi} \delta_p)^{n/2}}  \enspace, \label{musing}
\end{equation}
where we have used the value of $b_p$ from Eq.~\eqref{eq:wp},
$n$ is the number of space dimensions, and $M=\langle\rho_a\rangle L^n=\langle\rho_a\rangle (2\pi/\Delta)^n$ is the total axion mass in the volume.
The growth factor becomes
suppressed by the axion amplitude in the relevant momentum bin, and
scales with $1/\delta_p^{n/2}$.  Thus, in this regime, our estimate
from Eq.~\eqref{eq:mu-bar} has to be replaced by $\sigma_{\rm sing}$. We
clearly see this transition in the lower panels of
Fig.~\ref{fig:floquet}: at large $\delta_p$, only a single momentum bin
contributes to the resonance.  In the coherent case, it corresponds to
the resonance frequency at $m_a/2$, while for the random case, it can be
shifted away from $m_a/2$ due to the random phases of the axion.

We can model the transition between the two regimes by
introducing an exponential cut-off and patching the two regimes
together:
\begin{equation}\label{eq:mu_est}
  \sigma_{\rm eff} = \sigma e^{-\delta_p^2/\sigma^2} +
  \left(1 - e^{-\delta_p^2/\sigma^2}\right) \sigma_{\rm sing}  \enspace.
\end{equation}
This expression is shown by the black curves in the upper panels of
Fig.~\ref{fig:floquet}; the dashed curves correspond only to the first
term of the r.h.s.\ of Eq.~\eqref{eq:mu_est}. This simple model describes the coherent
clump pretty well, whereas some deviations appear for the random
phases, while the qualitative behavior is still captured. The generic
feature is the transition between the two regimes, set by the ratio
$\delta_p^2/\sigma^2$, which corresponds to the ratio of the
size of the clump (or axion coherence length) to the resonance length
(defined as the characteristic scale over which the parametric
resonance builds up). This has been pointed out by several authors in
the past \cite{Tkachev:1986tr, Tkachev:1987cd, Hertzberg:2010yz, Tkachev:2014dpa, Hertzberg:2018zte, Arza:2018dcy, Wang:2020zur}.

Let us comment briefly on the discretization related to the finite volume.
First, we note that in the ``large clump'' regime, $\delta_p \ll \sigma$,
the growth factor is independent of $\Delta$, see Eq.~\eqref {eq:mu-bar}.
In contrast, in the singular regime, $\delta_p \gg \sigma$, the growth factor is proportional to $\Delta^n$ for a clump with constant total mass, or to $\Delta^{n/2}$ for the random field with constant mean density. Therefore, the growth factor goes to zero in the continuum limit $\Delta\to 0$. However, for extended photon configurations filling the entire considered volume, the Fourier coefficients depend on the box size. In particular, for a homogeneous photon mode, the Fourier coefficients $A_{\vec k}$ are proportional to the box volume, and therefore the product $A_{\vec k} \sigma_\text{sing}$ becomes independent of $\Delta$ for a coherent clump. This can lead to physical effects which we are going to encounter in the following, and we call ``steady states''.

\section{Numerical simulations}\label{sec:num}

In this Section, we introduce the framework for our numerical simulations of photons in an axion background. We first address the problem of the axion momentum distribution, using  a WKB-like ansatz, similar to Method~A in the previous Section. The simulation will offer further insight into the phenomenology. In the next Section, we are going to apply the numerical framework to the case of a gravitational potential.

We start from Eq.~\eqref{eq1}, setting $A_0 = 0$, and ignoring axion gradients.
We simplify the problem by assuming that the electromagnetic wave propagates in the $z$ direction: %
\begin{eqnarray}
\left(\partial_t^2-\partial_z^2\right)A_x &=&  \gag \partial_ta\partial_zA_y \label{GeqAx1}
\\
\left(\partial_t^2-\partial_z^2\right)A_y &=&  -\gag \partial_ta\partial_zA_x \enspace. \label{GeqAy1}
\end{eqnarray}
For the axion background field, we assume 
\bb
a(t, z) = \frac{\sqrt{2\rho_a}}{m_a} \frac{1}{2i}
\left[ f(z)\ e^{i m_a t} - f^*(z)\ e^{-i m_a t} \right] \enspace.\label{axion_clump}
\ee
In the case of a coherent axion clump, $\rho_a$ is the maximum axion energy density and $f(z)$ is a peaked function whose maximum value is 1, while in the case of a random axion field, $\rho_a$ is the average axion energy density and $f(z)$ is normalized as
\bq
\frac{1}{L}\int_{-L/2}^{L/2} dz\, |f(z)|^2 =1 \enspace.
\eq 
We adopt a WKB ansatz for the photon,
\bq
A_x(t, z) &=& \alpha(t,z)\,e^{ik(z-t)} \nonumber
\\
A_y(t, z) &=& \beta(t,z)\,e^{ik(z+t)} \enspace,\label{photon}
\eq
where $k >0$ and $\alpha, \beta$ are slowly varying in both space and time.
The physical photon field is the real part of Eqs.~\eqref{photon}. 
We have switched again to linear polarization, $\alpha$ ($\beta$) corresponds to the amplitude of a wave traveling to the right (left).  
Using this ansatz and (\ref{axion_clump}) for the axion field, Eqs.~\eqref{GeqAx1} and~\eqref{GeqAy1} become, to leading order in the WKB approximation,
\bq
(\partial_t + \partial_z) \alpha 
&=&  -\sigma f^*(z)\, \beta\,e^{i\varepsilon t}\label{analogousalpha}
\\
(\partial_t - \partial_z) \beta 
&=& -\sigma f(z) \,\alpha\,e^{-i\varepsilon t}\label{analogousbeta} \enspace,
\eq
where we have neglected terms that cannot resonate, 
$\sigma$ has been defined in Eq.~\eqref{eq:mu-bar},
and, in analogy to $\epsilon_{\vec k}$ from Eq.~\eqref{eeps1},
we have introduced the quantity
\bq
\varepsilon \equiv 2k - m_a \enspace. \label{eq:vareps}
\eq
Notice that $\varepsilon$ differs from $\epsilon_{\vec{k}}$ as it does not include the axion momentum.
We Fourier expand the slowly varying photon amplitudes as
\bb
\alpha(t, z) = \int_{-\infty}^{\infty} \frac{dr}{2\pi}\, \alpha_r(t)\ e^{irz} \enspace, \qquad
\beta(t, z) = \int_{-\infty}^{\infty} \frac{dr}{2\pi}\, \beta_r(t)\ e^{irz}  \enspace.
\ee
We choose $\sigma$ as our unit energy and define 
\bb
\tau = \sigma t \,,\qquad \zeta = \sigma z \,,\qquad \tilde{r} =\frac{r}{\sigma} \,,\qquad \rm etc.
\ee
In the following, the tilde denotes quantities in units of $\sigma$.
After redefining the amplitudes as
$\alpha_{r} \to e^{i \tilde{\varepsilon}\tau/2} \alpha_{r}$, 
$\beta_{r}  \to e^{-i \tilde{\varepsilon}\tau/2} \beta_{r}$,  
the equations for the slowly varying photon amplitudes become
\bq
\partial_\tau \alpha_r 
&=&  
- i\left(\frac{\tilde{\varepsilon}}{2} +\tilde{r} \right) \alpha_r
-\int_{-\infty}^{\infty} \frac{d\tilde{s}}{2\pi}  \beta_s\ F^*(-\tilde{r}+{\tilde{s} })
\label{eqalpha}\\
\partial_\tau \beta_r
&=& 
 i\left(\frac{\tilde{\varepsilon}}{2} +\tilde{r}  \right) \beta_r
-\int_{-\infty}^{\infty} \frac{d\tilde{s}}{2\pi} \alpha_s\ F(\tilde{r}-{\tilde{s} })\enspace.\label{eqbeta}
\eq
where 
\bb
F( \tilde{u}) = \int_{-\infty}^{\infty} d\zeta\ 
\ f(\zeta)\ e^{-i  \tilde{u}\zeta }\enspace.
\ee
We solve Eqs.~(\ref{eqalpha},~\ref{eqbeta}) numerically by 
placing our system in a one dimensional box of length $\tilde{L}$
with $N$ grid points. The space resolution is $\Delta_\zeta = \tilde{L}/N$, while the momentum resolution is $\Delta = 2\pi/\tilde{L}$. 
For details regarding the implementation, see Appendix~\ref{num_detail}.

\subsection{Coherent Gaussian clump}
\label{sec:coh_clump}

\begin{figure}[t!]
  \centering
  \includegraphics[width=0.8\linewidth]{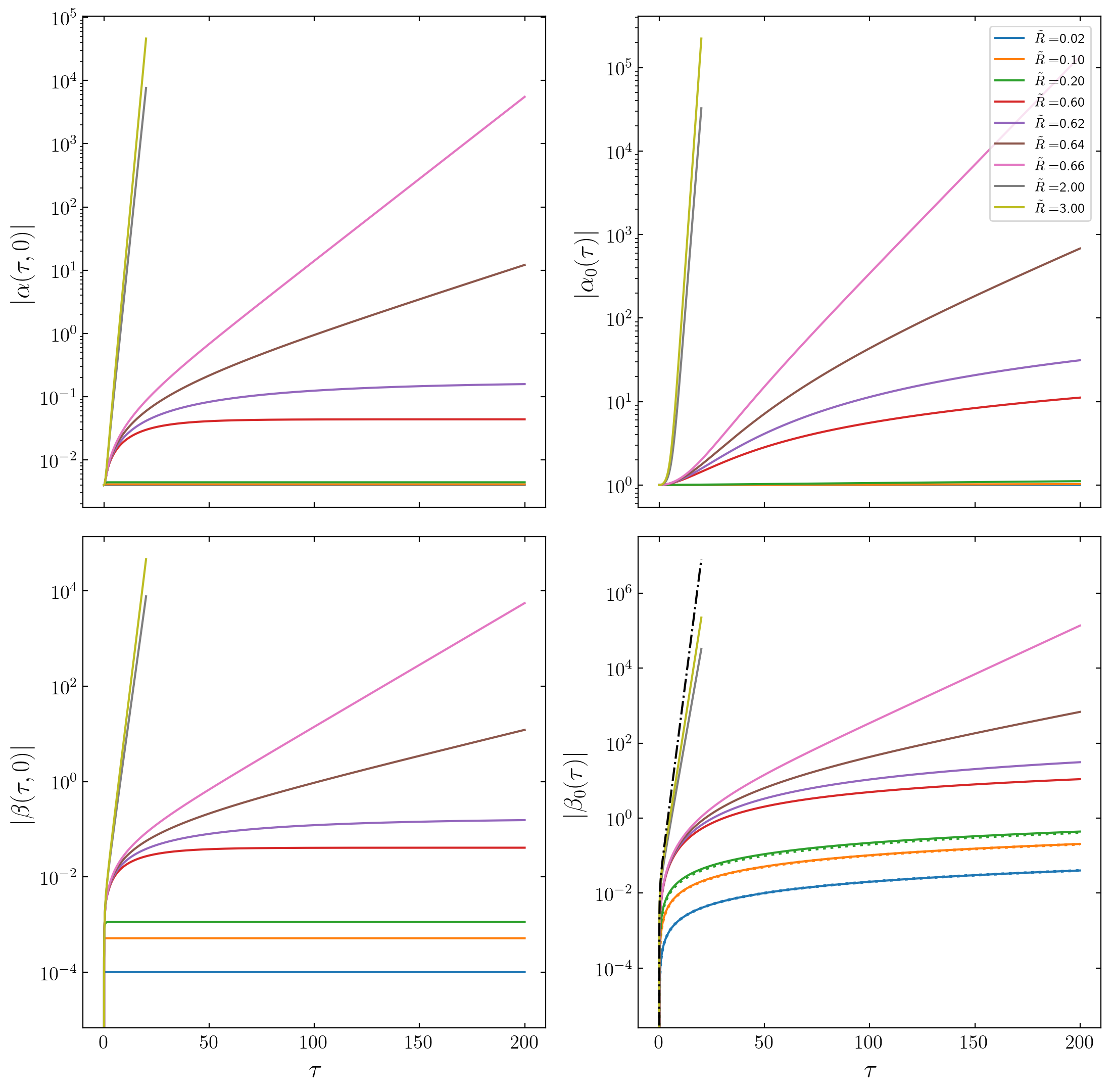}
	\caption{The left panels show the photon amplitudes at the center of the axion clump $\zeta=0$ as a function of time for different clump widths $\tilde R$, while the right panels show the time evolution of the homogeneous mode $\tilde{k}=0$. For all lines, the simulation parameters are $\tilde{L}=250$, $\tilde{\varepsilon}=0$, $\alpha_k(0) =\delta_{k0}$, $\beta_k(0)=0$. 
	From the left panels, we see the transition between exponential growth and  saturation of the growth around $\tilde{R} = 0.63$.
	In the right lower panel, the dotted lines represent the prediction from Eq.~\eqref{betasing}.
	The dash-dotted lines represent Eq.~\eqref{betahomo}, which we divided by 30 to bring  the dash-dot line close to the solid lines.
	}\label{fig:centers} 
\end{figure}%

We consider first a Gaussian axion clump of width $\tilde{R}$ in coordinate space:
\bq
f(\zeta) = \e^{-\frac{1}{2} \left(\frac{\zeta}{\tilde{R}}\right)^2 } \enspace. \label{Gclump}
\eq
In the notation of Section~\ref{sec:dispersion}, we have $\tilde{R}=\sigma/\delta_p$.
From Eq.~\eqref{eq:mu_est}, we know that, at $\tilde{R} \sim 1$, there is a transition between a regime of exponential growth (large clump) and a regime where the exponential growth cannot develop (small clump). We test this result numerically.

The left panels of Figure~\ref{fig:centers} show the behavior of the photon amplitudes at the center of the axion clump as a function of time for several values of $\tilde{R}$, while the right panels show the time evolution of the Fourier mode $\tilde{k}=0$, corresponding to the resonant frequency, i.e., the homogeneous component. We choose the initial condition for the photon field to be $\alpha_k(0) =\delta_{k0}$, $\beta_k(0)=0$ and $\tilde{\varepsilon}=0$. From the left panels, we see that, for $\tilde{R}\lesssim 0.63$,  after a period of initial growth, the amplitudes at the center of the clump reach a constant value. On the other hand, for $\tilde{R}\gtrsim 0.63$, the amplitudes keep growing exponentially.

The right panels show that $\beta_0(\tau)$ grows more than $\alpha_0(\tau)$, implying that $\beta_0(\tau)$ has a greater overlap with the eigenvector corresponding to the maximum growth factor.
We can then get a quantitative match with the predictions of the previous Section by looking at the lower right panel. Well into the singular regime ($\tilde{R} = 0.02, 0.10, 0.20$), we get excellent agreement between Eq.~\eqref{eq:mu_est} and the simulations. In fact, in this regime $\beta_0(\tau)$
essentially coincides with the eigenvector with maximum growth factor.
The dotted lines represent our prediction and are given by
\bq
\beta_0(\tau) = \alpha_0(0)\sinh(\tilde{\sigma}_{\rm sing} \tau) \enspace.
\label{betasing}
\eq 
They superimpose extremely well with the numerical result. 
Eq.~\eqref{betasing} is the expected solution for $\beta_0(\tau)$ given our initial condition for the photon, Eq.~\eqref{eq:mu_est} and Eq.~\eqref{solbk1}.

In the intermediate regime ($\tilde{R} = 0.60, 0.62, 0.64, 0.66$), and in the full exponential regime ($\tilde{R} = 2.00, 3.00$), the homogenous mode does not coincide with the eigenvector that grows the fastest, and thus we do not expect the lines on the right panels to grow at the rate given by Eq.~\eqref{eq:mu_est}. However, we can check that in the limit $\tilde{R}\gg \tilde{L}$, the lines converge to the result for a homogeneous axion field, given by
\bq
\beta_0(\tau)=\alpha_0(0)\sinh(\tau) \enspace. \label{betahomo}
\eq 
The black dash-dotted line represents the expression above, up to a re-scaling by a constant factor that we introduced to allow an easy comparison of the slope with the other lines. We see that indeed the slope converges to a constant value as $\tilde{R}$ increases. 

\begin{figure}[t!]
  \centering
  \includegraphics[width=0.8\linewidth ]{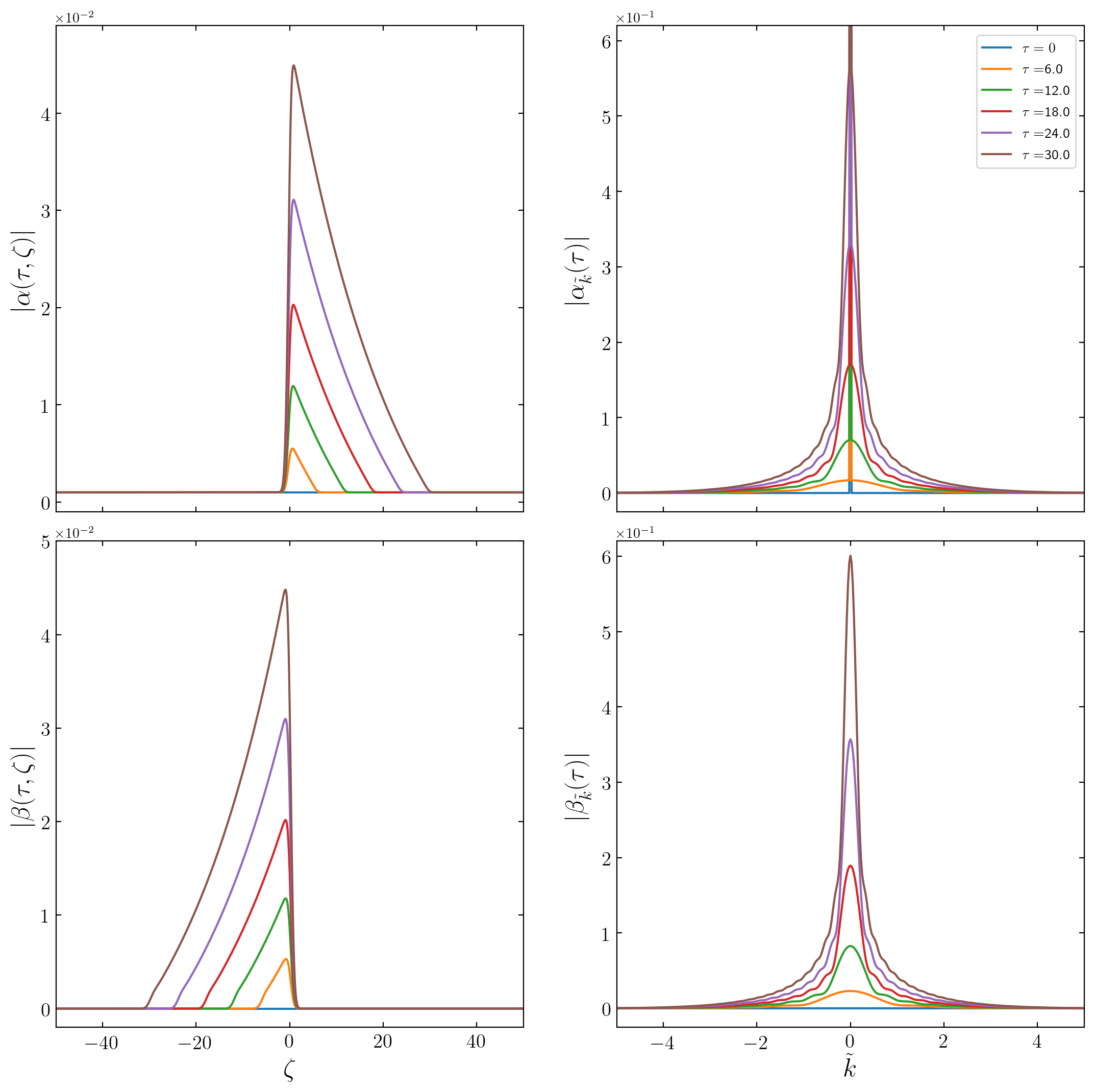}
	\caption{Output of the simulation for a Gaussian axion clump, Eq.~\eqref{Gclump}, with width $\tilde{R}=0.65$. 
		The left panels show the result for the two photon polarizations as a function of the space coordinate $\zeta$, while the right panels show the result in momentum space. The colored lines represent different times between $\tau=0$ and $\tau=30$. The initial condition for the photon is $\alpha_k(0) =\delta_{k0}$, $\beta_k(0)=0$.
	The other parameters of the simulation are: $\tilde{L}=1000$, $\tilde{\varepsilon}=0$. The  height of the upper right panel has been limited in order to show the evolution of the non-zero modes better; the value of $\alpha_0(\tau)$ is always larger than 1.
    The clump size $\tilde{R}=0.65$ is just in the regime of exponential growth (``large clump''), c.f.~Fig.~\ref{fig:centers}.
	The width of the resonance band is $2\sigma$. We see from the right panels that the photon modes within the band grow in time.
	In physical space, the photon amplitudes keep growing in time as they propagate.
	\label{fig:Gexp} }
\end{figure}%

\begin{figure}[t!]
  \centering
  \includegraphics[width=0.8\linewidth]{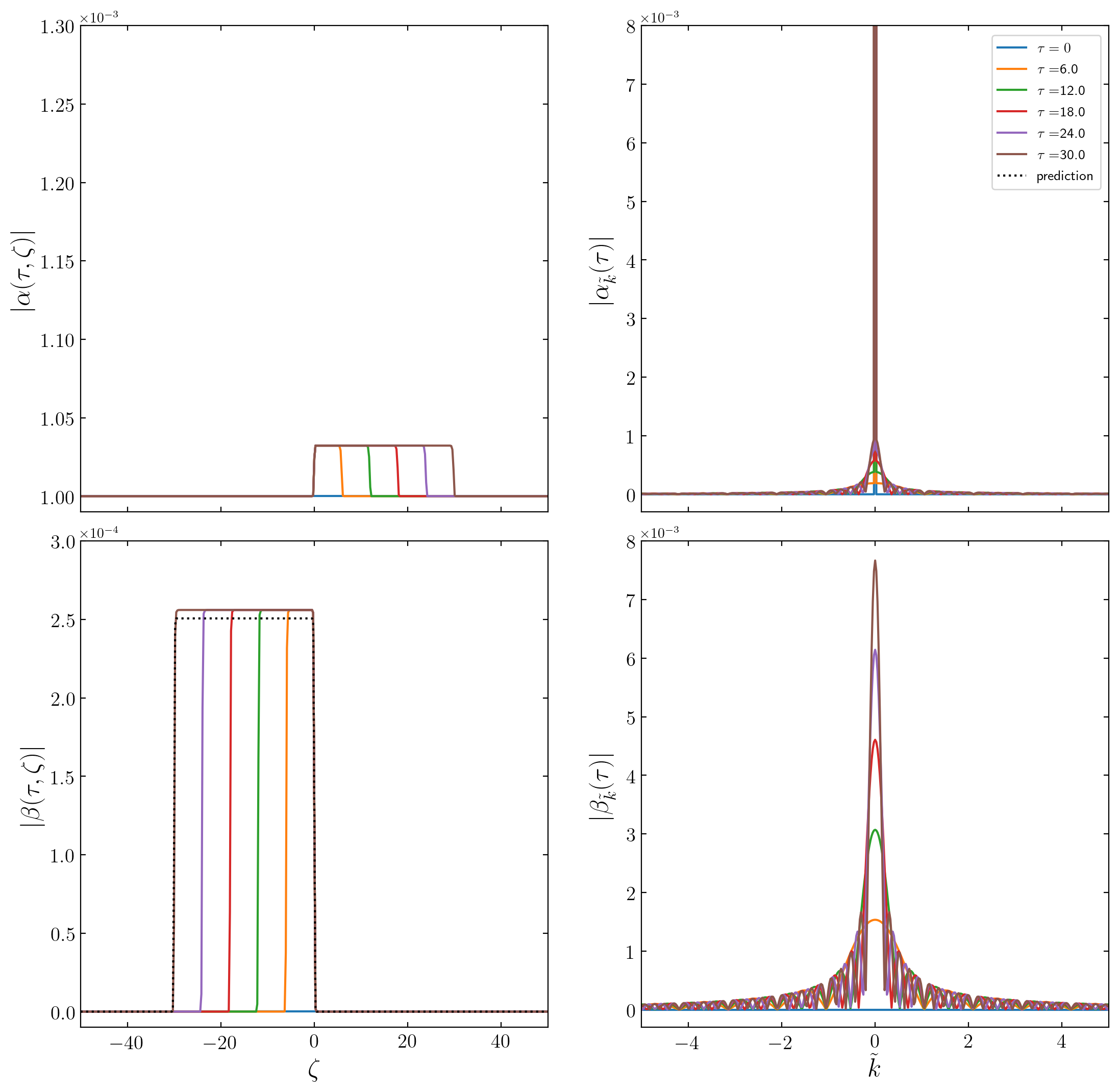}
	\caption{Output of the simulation for a Gaussian axion clump, Eq.~\eqref{Gclump}, with width $\tilde{R}=0.1$.
	All other parameters are the same as those of Fig.~\ref{fig:Gexp}.
	We are in the regime where the matrix Eq.~\eqref{eq:beta3} is singular (``small clump''). From the lower right panel, we see that the homogeneous photon amplitude, corresponding to the resonant frequency, grows, while the amplitudes oscillate in time.
	In physical space, this situation manifests itself as the appearance of a wavefront of constant intensity that propagates. The black dotted line in the lower left panel represents the prediction Eq.~\eqref{ss1solbeta12} at $\tau = 30$. \label{fig:Gsteady} }
\end{figure}%

Figures~\ref{fig:Gexp} and~\ref{fig:Gsteady} show the photon amplitudes in real (left panels) and momentum (right panels) space at six different times equally spaced between $\tau=0$ and $\tau=30$, in the exponential and in the singular regime, respectively.
In Fig.~\ref{fig:Gexp}, the width of the Gaussian axion clump is $\tilde{R} = 0.65$. We see that $\alpha(\tau,\zeta)$ and $\beta(\tau,\zeta)$ grow exponentially at the location of the axion clump. Moreover, the two photon polarizations propagate in opposite directions. Given enough time, the electromagnetic wave would fill the whole box. Correspondingly, we see that $\beta_0(\tau)$ is the fastest-growing Fourier mode ($\alpha_0(\tau)$ is already large at $\tau=0$ due to our choice of initial conditions). From the right panels, we see that the width of the resonance band is approximately $2\sigma$, in accordance with the expectation from Section~\ref{sec:dispersion}.

In Fig.~\ref{fig:Gsteady}, the width of the Gaussian axion clump is $\tilde{R} = 0.1$, corresponding to the small clump regime. The maximum values of the amplitudes $\alpha(\tau, \zeta)$ and $\beta(\tau, \zeta)$ do not grow exponentially, but rather stay constant in time. As the waves propagate, the homogeneous Fourier mode grows, while other modes oscillate. In coordinate space, this corresponds to the appearance of a wavefront of constant intensity that propagates. We call this phenomenon ``steady state''. As mentioned at the end of Section~\ref{sec:dispersion}, this effect appears in case of photon initial conditions extending till the boundary of the considered volume, c.f.~left-upper panel of Fig.~\ref{fig:Gsteady}.

As there is no exponential growth in the ``small clump" regime, we can solve Eqs.~(\ref{analogousalpha}) and (\ref{analogousbeta}) perturbatively and provide an analytical expression for the steady state. We consider $\varepsilon=0$ and write $\alpha=\alpha^{(0)}+\alpha^{(2)}+...$ and $\beta=\beta^{(1)}+\beta^{(3)}+...$, where the zeroth order corresponds to a homogeneous initial photon amplitude $\alpha^{(0)}(t,z)={\cal A}\Theta(t)$. Using the dimensionless parameters defined above, $\beta^{(1)}$ can be found solving the equation
\bb
(\partial_\tau-\partial_\zeta)\beta^{(1)}=-{\cal A}\Theta(\tau)\,e^{-{1\over2}{\zeta^2\over\tilde R^2}}. \label{ss1eqbeta11}
\ee
The solution to this equation is derived in Appendix~\ref{steady_detail}. The result is
\bb
\beta^{(1)}(\tau,\zeta)=-{\cal A}\sqrt{\pi\over2}\tilde R\,\left[\text{erf}\left({\tau+\zeta\over\sqrt{2}\tilde R}\right)-\text{erf}\left({\zeta\over\sqrt{2}\tilde R}\right)\right]. \label{ss1solbeta12}
\ee
For $\tau\gg \tilde R$, in other words times much longer than the crossing time of the clump, $t \gg R$,
the growth of $\beta^{(1)}$ saturates, reaching a steady state given by
\bb
\beta_\text{steady}^{(1)}(\zeta)=-{\cal A}\sqrt{\pi\over2}\tilde R\,\left[1-\text{erf}\left({\zeta\over\sqrt{2}\tilde R}\right)\right]. \label{ss1solbeta13}
\ee
Equation~\eqref{ss1solbeta12} is shown to match the numerical simulations in Fig.~\ref{fig:Gsteady}. The small discrepancy in the height of the wavefront is due to the fact that Eq.~\eqref{ss1solbeta12} is obtained using a perturbative approach. The discrepancy is reduced if $\tilde{R}$ is made smaller. From Eq.~\eqref{ss1solbeta13}, in the limit $\zeta\rightarrow -\infty$ ($\beta$ propagates in the negative $z$ direction), we see that the height of the wave front goes to zero proportionally to $\tilde{R}$ for $\tilde{R}\rightarrow 0$.

\subsection{Random phases}

\begin{figure}[t!]
  \centering
  \includegraphics[width=0.8\linewidth]{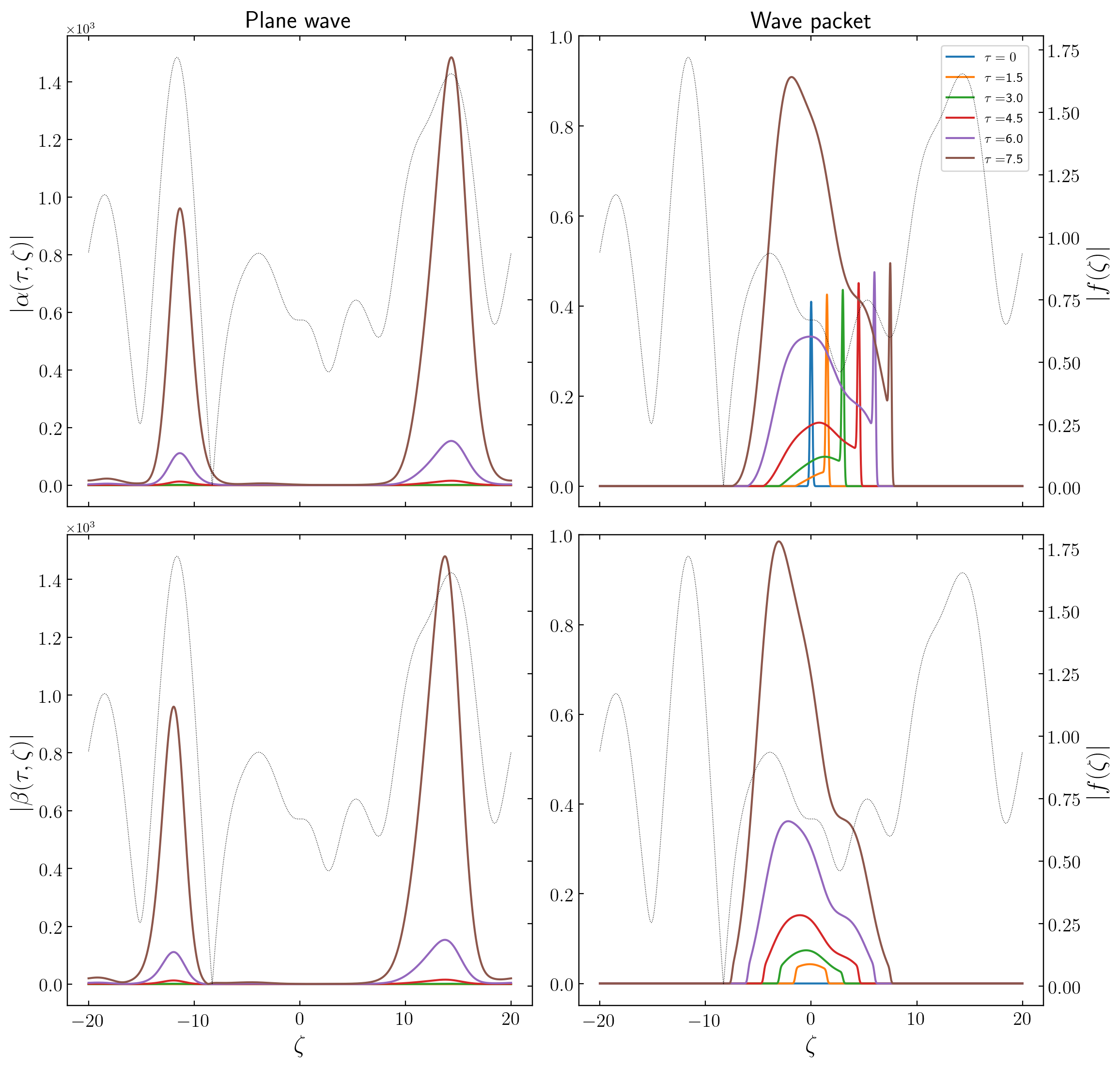}
	\caption{Output of the simulation for a random axion field with coherence length $\tilde{R}=2$. The black dotted line represent the axion profile $|f(\zeta)|$.
	For the left panels, the initial condition for the photon is $\alpha_k(0) =\delta_{k0}$, $\beta_k(0)=0$, while, for the right panels, it is a Gaussian wave packet with width 0.1, see Eq.~\eqref{wp}.
	The other parameters of the simulation are: $\tilde{L}=40$, $\tilde{\varepsilon}=0$. 
	}\label{fig:random} 
\end{figure}%

Finally, we test our predictions from Section~\ref{sec:dispersion} against the simulations for the case of a random axion field, whose Fourier amplitudes are defined according to Eq.~\eqref{eq:rho_m_gauss}. Figure~\ref{fig:random} shows the behavior for an axion background field with coherence length $\tilde{R}=2$. The left panels show the result when the photon is initialized to be a plane wave $\alpha_k(0) =\delta_{k0}$, $\beta_k(0)=0$, while, for the right panels, the photon is initially a wave packet
\bq
\alpha(0, \zeta) = 0.4\  e^{-\frac{1}{2}\left(\frac{\zeta}{0.1}\right)^2} \qquad \beta(0, \zeta)=0 \enspace. \label{wp}
\eq
Since the size of the coherent patches is larger than the resonance length, the photon amplitudes grow exponentially within each patch. 
This is clearly visible on the left panels, where we see that at time $\tau =7.5$, the amplitudes have reached values of order $10^3$ at the locations where the axion density is the largest (the initial value of $\alpha(\tau, \zeta)$ is 0.025). Notice that the rightmost peak is higher than the left one, although $f(\zeta)$ has approximately the same value at either location. This is due to the rightmost peak being broader.
From the top right panel, we see the initial wave packet propagating to the right, and a wake appearing with a shape resembling that of $|f(\zeta)|$. The $\beta(\tau, \zeta)$ amplitude also has a similar shape.

We have also tested the singular regime by considering an random axion field with coherence length down to $\tilde{R}=0.01$. The results from the numerical simulations are in excellent agreement with the prediction Eq.~\eqref{betasing}.

\section{Gravitational redshift}\label{sec:redshift}

In this Section, we study the parametric resonance considering that the emission of photons takes place in the presence of a gravitational potential. We focus on static potentials from astrophysical objects, for instance, galaxies or other gravitationally bound systems. Although we have in mind that the axions provide the dark matter, and we take dark matter halos as benchmark configurations to determine the axion amplitude, we do not require a self-consistent solution of the axion field equation coupled to gravity. Instead, we impose an external gravitational potential with properties motivated by dark matter halos.

For a curved space-time with metric tensor $g_{\mu\nu}$, the covariant form of (\ref{eqcov1}) and (\ref{eqcov2}) is
\begin{eqnarray}
\partial_\mu\left(\sqrt{-g}g^{\mu\rho}g^{\nu\sigma}F_{\rho\sigma}\right) &=& \sqrt{-g}\, {1\over2} \gag \,\epsilon^{\mu\nu\rho\sigma}\partial_\mu a F_{\rho\sigma} \label{GeqAcov}
\\
\partial_\mu\left(\sqrt{-g}g^{\mu\nu}\partial_\nu a\right)+\sqrt{-g}m_a^2a &=&
\frac{1}{\sqrt{-g}}\,{1\over8} \gag \,\epsilon^{\mu\nu\rho\sigma}F_{\mu\nu}F_{\rho\sigma}\enspace, \label{Geqavoc}
\end{eqnarray}
where $g$ is the determinant of the metric, $\epsilon^{\mu\nu\rho\sigma}$ is the Levi-Civita symbol and $\epsilon^{0123}=1$.
We work in the weak gravity limit, so that the metric takes the form
\bq
ds^2 = (1 + 2\Phi) dt^2 - (1 - 2\Phi) \delta_{ij} dx^i dx^j  \enspace,\label{Gmetric}
\eq
where $\Phi(\vec x)$ is the time independent Newtonian gravitational potential
with $\Phi(\vec x)<0$ and $|\Phi(\vec x)| \ll 1$.
We adopt the Coulomb gauge $\delta^{ij}\partial_iA_j=0$ and neglect terms proportional to $\Phi^2$, $\partial_i\Phi$,  $\gag\partial_i a$. Requiring $A_0$ to vanish at spatial infinity, we set $A_0 = 0$ and obtain
\bq
\left(\partial_0^2 - (1+4\Phi) \nabla^2   \right) A^i
&=& \gag \partial_0 a\, \epsilon^{\,0ijk} \partial_j A_k \label{thee}\\
(1-2\Phi) \partial_0^2 a - (1+2\Phi) \nabla^2  a + m_a^2 a &=& 0\enspace, \label{eom_axion}
\eq
with $\nabla^2 = \delta^{kl} \partial_k\partial_l$. Again, we are not considering back-reactions on the axion field.
In deriving these equations, we have assumed that gradients of the potential are small and can be neglected compared to all other relevant terms in the equations of motion, consistent with Eqs.~(\ref{eq:approx1}, \ref{eq:approx}) below. However, as we will discuss, we do take into account potential gradients in the phase of the axion field, where they will eventually lead to the detuning of the resonance, c.f., Eqs.~(\ref{detune}, \ref{eq:cond-pot}).

We restrict ourselves to the one-dimensional case, in which electromagnetic and axion fields propagate along the same direction $\hat z$, which is also the direction of the gradient of $\Phi$. The equations of motion become
\begin{eqnarray}
\left(\partial_t^2-\partial_z^2\right)A_x &=&  \gag (\partial_ta)\partial_zA_y \label{GeqAx2}
\\
\left(\partial_t^2-\partial_z^2\right)A_y &=& - \gag (\partial_ta)\partial_zA_x \label{GeqAy2}
\\
\left(\partial_t^2-\partial_z^2+m_a^2(1+2\Phi)\right)a &=& 0 \enspace. \label{geqa1}
\end{eqnarray}
To obtain Eqs.~(\ref{GeqAx2}), (\ref{GeqAy2}) and (\ref{geqa1}), we have redefined $dz\rightarrow dz/(1-2\Phi)$ and neglected terms of order $\gag\Phi$. We can already see that the main effect of the gravitational potential is to modify the dispersion relation for the axion field. With this choice of coordinates, the photon equations of motion (\ref{GeqAx2}, \ref{GeqAy2}) have the same form as in flat space, c.f.~Eqs.~(\ref{GeqAx1}, \ref{GeqAy1}). As we show in Appendix~\ref{eikonal}, the relative frequencies of the axion and photon fields are not affected by the redshift.\footnote{This statement does not depend on the particular choice of coordinates adopted in Eqs.~(\ref{GeqAx2}, \ref{GeqAy2}, \ref{geqa1}). It follows from the solution of the geodesic equations in the eikonal approximation, which is equivalent to solving Eqs.~\eqref{thee} and \eqref{eom_axion} in the relevant limit. Here we do not agree with the discussion in Ref.~\cite{Wang:2020zur}.} It is the impact of the axion three-momentum that leads to the detuning of the resonance discussed in the following.

\subsection{The axion field in the presence of a gravitational potential}

As we have seen in Secs.~\ref{sec:dispersion} and \ref{sec:num}, the resonance can only develop if the momentum spread of the axions is much less than the growth factor. Hence, in order to study the effect of the gravitational redshift, in this Section, we require this condition to be fulfilled, i.e., we assume a very cold axion distribution, such as, for instance, a condensate. We consider two specific examples for such a scenario: 
\begin{itemize}
\item[(a)] 
a monochromatic axion wave with a wave vector $\vec{p}_*$ at infinity (similar to the case considered in Section~\ref{sec:essentials}), which ``falls'' into the gravitational potential well along a geodesic;  
\item[(b)]
an axion wave which is bound inside the gravitational potential and corresponds to a stationary state with fixed energy $E$.
\end{itemize}
In both cases, the effect of the gravitational potential is determined by 
Eq.~(\ref{geqa1}) and, as we show in Appendices~\ref{eikonal} and \ref{wkb}, the solution to that equation can be written as
\bb
a(t,z)={\sqrt{2\rho_a}\over m_a}\sin\left(m_a t-S(z)\right) \enspace, \label{Gsola1}
\ee
where
\bb
S(z)=\int_{z_*}^zdz'\sqrt{p^2_*-2m_a^2\Phi(z')} \enspace. \label{GS1}
\ee
In Eq.~\eqref{GS1}, $p_*$ is a constant with $|p_*| \ll m_a$. In case (a), $p_*$ is the $z$ component of the covariant axion four-momentum $p_\mu$ evaluated at a far away location $z_*$ where the potential is zero, and in case (b), $p_*^2 = 2m_aE$ with $E < 0$ corresponding to the binding energy inside the potential. Eqs.~\eqref{Gsola1} and \eqref{GS1} are obtained in the eikonal approximation in case~(a) and the WKB approximation in case~(b), see Appendices~\ref{eikonal} and \ref{wkb} for more details. In both cases, the approximation is that the potential changes slowly on the scale of the deBroglie wavelength $\lambda$ of the axion:
\bq
\frac{\lambda}{2\pi} \equiv \frac{1}{\partial_z S(z)} = 
\frac{1}{\sqrt{p^2_*-2m_a^2\Phi}} \ll
\frac{p_*^2/m_a^2 - 2 \Phi}{|\partial_z\Phi|}
\label{eq:approx1}
\eq
or, assuming $|p_*|/m_a \ll \sqrt{-2\Phi}$,
\bq
\left|\frac{\partial_z\Phi}{\Phi}\right|
\ll m_a\sqrt{-\Phi} 
\enspace.
\label{eq:approx}
\eq
Let us consider, as an example, the gravitational potential from the dark matter in our galaxy around the location of the Sun.\footnote{For a more accurate calculation a realistic model of the Milky Way would have to be considered, taking into account also the baryonic contributions to the potential.} We assume an NFW profile, with the potential $\Phi(z) = -4\pi G_N\rho_s r_s^2 \log(1+z/r_s)/(z/r_s)$, 
adopting typical values for the parameters $\rho_s$ and $r_s$ (see e.g., \cite{Wang:2015ala}). We find 
\bq
\Phi \sim 10^{-6} \,,\qquad
\partial_z\Phi \sim 10^{-34} \,{\rm eV} \enspace.
\label{eq:pot}
\eq
We see that for relevant values of $m_a$ the condition \eqref{eq:approx} is fulfilled to a very good accuracy.

From Eq.~\eqref{GS1}, it follows that the gravitational potential becomes important for $|p_*|/m_a \lesssim \sqrt{-\Phi}$. Considering the typical values from Eq.~\eqref{eq:pot}, this amounts to axion velocities of less than $10^{-3}$ in case (a), which we are going to assume in the following. For case (b), we assume that the binding energy $E$ fulfills $|E| \ll m_a |\Phi|$, see App.~\ref{wkb} for justifications and further comments. For numerical estimates, we will work under the assumption $|p_*|/m_a \ll \sqrt{-2\Phi}$. Finally, let us stress that here we are not concerned with the question of whether those are realistic assumptions for an axion dark matter halo in a galaxy. The purpose of the analysis is to study under which conditions an enhancement of the photon flux can develop.

\subsection{Estimate of the detuning length}

We consider a region of space around a point $\Bar{z}$ where the gravitational potential can be approximated as a constant. Then, the axion field of Eqs.~\eqref{Gsola1} and \eqref{GS1} resembles a monochromatic axion, as in Eq.~\eqref{axionfield1}, with an effective momentum $p_{\rm eff} = \sqrt{p_*^2-2m_a^2\Phi(\Bar z)}$. We can then use the equations of Sec.~\ref{sec:essentials}, in particular, the resonance band is given by Eq.~\eqref{band}. 

\begin{figure}[t!]
  \centering
  \includegraphics[width=0.9\linewidth]{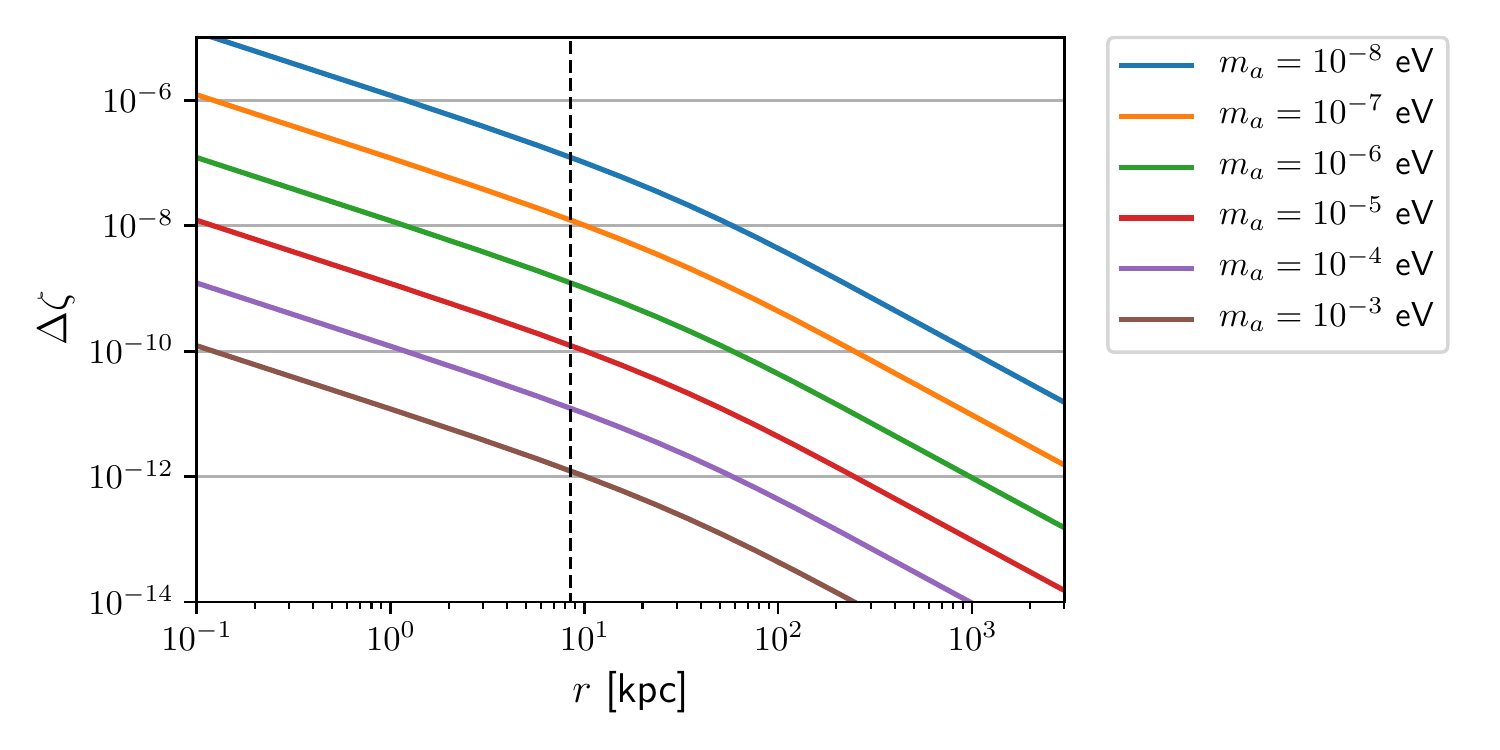}
	\caption{Detuning length over resonance length, $\Delta\zeta$, as a function of the location in our galaxy for different axion masses, assuming that the axion density, as well as the gravitational potential, are set by a typical NFW dark matter profile with parameters taken from \cite{Wang:2015ala}. The axion--photon coupling is set to $\gag = 10^{-11}$~GeV$^{-1}$. The dashed line indicates the location of the Sun.
	}\label{fig:delta_zeta} 
\end{figure}

Consider a photon with momentum $k$. Depending on the location $\Bar z$, it will feel a different effective axion momentum, and hence it may be shifted out of the resonance band, depending on the value of $p_{\rm eff}(\Bar z)$.
The distance required to detune a photon from the resonance can be estimated considering the boundaries of the resonance band $-2\sigma < \epsilon_{\vec k} < 2\sigma$. We define 
two locations $z_1$ and $z_2$ such that
\bq
2 k  - m_a - \sqrt{p^2_* - 2m_a^2\Phi_1}&=& - 2\sigma \enspace, \nonumber\\
2 k  - m_a - \sqrt{p^2_* - 2m_a^2\Phi_2 }&=& + 2\sigma\enspace, \label{detu}
\eq
where we have used $\epsilon_{\vec k}$ from Eq.~\eqref{eeps1} with the substitution $p\to p_{\rm eff}$, and $\Phi_{1,2} \equiv \Phi(z_{1,2})$.
Hence, a photon with frequency $k$ enters the resonance band at $z_1$ and exits it at $z_2$, or viceversa. From Eq.~\eqref{detu}, we obtain the following estimate for the
detuning distance in units of the resonance length $\sigma^{-1}$:
\bq
\Delta\zeta \equiv \sigma |z_1-z_2|
\approx \frac{4\sigma^2}{m_a^2}\frac{ \sqrt{p^2_* - 2m_a^2 \Phi}}{|\partial_z\Phi|} \enspace.\label{detune}
\eq
Setting $|p_*|/m_a \ll \sqrt{-2\Phi}$, and
using the typical values for the potential for our galaxy at the Sun's location from Eq.~\eqref{eq:pot}, as well as $\sigma$ from Eq.~\eqref{sigma_value}, we obtain 
\bq
\Delta\zeta \sim 10^{-10} 
\left( \frac{10^{-5}~\mathrm{eV}}{m_a}\right)
\left( \frac{\gag}{10^{-11}~\mathrm{GeV}^{-1}} \right)^2
\left( \frac{\rho_a}{0.4~\mathrm{GeV/cm}^3}\right)\enspace.
\eq
In Fig.~\ref{fig:delta_zeta}, we show $\Delta\zeta$ as a function of the distance from the galactic center for different values of the axion mass.  The potential, its derivative, and the growth factor $\sigma$ are calculated at each location according to a typical
NFW profile for the dark matter, taken from \cite{Wang:2015ala}.

We conclude that the gravitational redshift detunes the photons on distances corresponding to tiny fractions of the resonance length at any location in our galaxy. This implies that no resonance can develop, even under the assumption of a completely cold axion condensate. This means, in particular, that the bounds derived in Ref.~\cite{Sigl:2019pmj} do not apply since the redshift effect has not been taken into account. 

In order for a resonance to develop, we need $\Delta\zeta \gtrsim 1$, which implies
\bq
\left|\frac{\partial_z\Phi}{\sqrt{-2\Phi}}\right| \lesssim
\frac{10^{-18}}{1\, \rm pc} 
\left( \frac{10^{-5}~\mathrm{eV}}{m_a}\right)
\left( \frac{\gag}{10^{-11}~\mathrm{GeV}^{-1}} \right)^2
\left( \frac{\rho_a}{0.4~\mathrm{GeV/cm}^3}\right)\enspace,
\label{eq:cond-pot}
\eq
requiring extremely high axion densities and/or extremely flat gravitational potentials. For comparison, for a typical NFW profile the quantity on the left-hand side is roughly of order $10^{-8}/\rm pc$ at the location of the Sun.

\subsection{Numerical results}

We present now the results of numerical simulations.
With Eq.~(\ref{Gsola1}) for the axion field, and using the ansatz Eq.~\eqref{photon} for the photon, Eqs.~(\ref{GeqAx2},~\ref{GeqAy2}) become
\bq
(\partial_t + \partial_z) \alpha 
&=& -\sigma \beta\,e^{i(\varepsilon t+S(z))} \label{eqalphaG}
\\
(\partial_t - \partial_z) \beta 
&=& -\sigma \alpha\,e^{-i(\varepsilon t+S(z))} \label{eqbetaG}\enspace.
\eq
We have neglected terms that cannot resonate, $\sigma$ is given in Eq.~\eqref{esigma1} and $\varepsilon$ in Eq.~\eqref{eq:vareps}.
Equations~(\ref{eqalphaG},~\ref{eqbetaG}) are analogous to Eqs.~(\ref{analogousalpha},~\ref{analogousbeta}). In order to study the system numerically, we consider a toy model, which nevertheless captures the relevant physics. We solve Eqs.~(\ref{eqalphaG},~\ref{eqbetaG})
using the code described in Section~\ref{sec:num} and Appendix~\ref{num_detail}, with
\bb
F( \tilde{u}) = \int_{-\infty}^{\infty} d\zeta\ 
\ e^{-i S(\zeta)}\ e^{-i  \tilde{u}\zeta } \enspace.
\ee
We place our system in a box of length $\tilde{L}=500$ with periodic boundary conditions. For our toy model, we choose a Gaussian gravitational potential
\bb
\Phi(\zeta) = -\frac{\Phi_0}{\sqrt{2\pi } \tilde{R}_\Phi}\ e^{-\frac{1}{2} \left(\frac{\zeta}{\tilde{R}_\Phi }\right)^2 } \enspace, \label{gpot}
\ee
with  $\tilde{R}_\Phi = \tilde{L}/16$ and $\Phi_0 = 0.3$.

\begin{figure}[t!]
  \centering
  \includegraphics[width=0.6\linewidth]{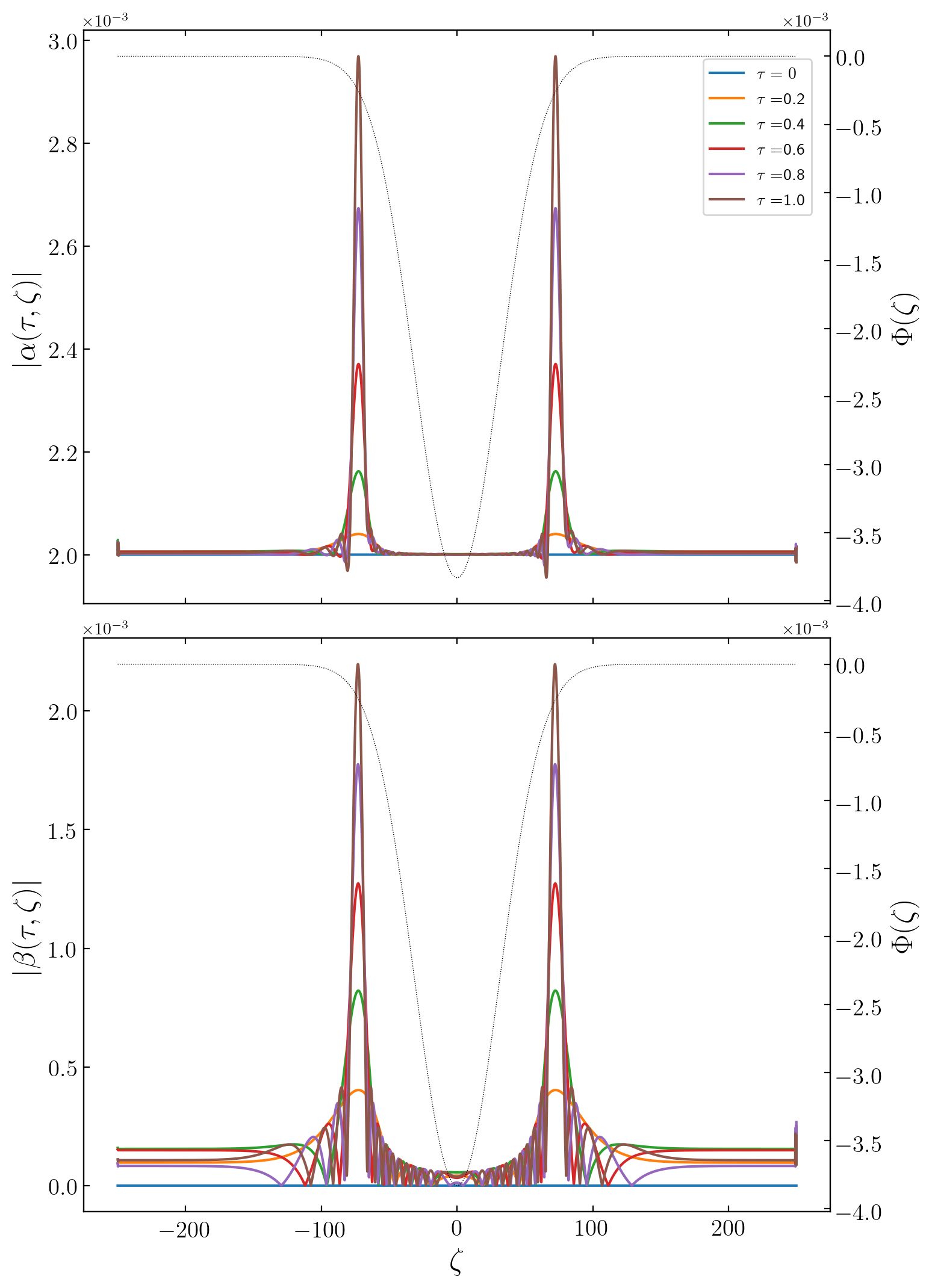}
	\caption{Photon amplitudes at different times with the gravitational potential Eq.~\eqref{gpot} and the background axion field given by Eq.~\eqref{Gsola1}. The parameters of the simulation are $\tilde{L}=500$, $\tilde{m}_a=1100$, $\tilde{p}_* = 0$, $\alpha_k(0) =\delta_{k0}$, $\beta_k(0)=0$. The parameter $\tilde{\varepsilon}$ is set to 25, such that the photon grows the most at $\zeta= \pm 72.6$. At this location the detuning distance is $\Delta \zeta = 4.3$, see Eq.~\eqref{detune}. The gravitational potential is represented by the dotted black line.} \label{fig:peaks}
\end{figure}%

Figure~\ref{fig:peaks} shows a typical result.
Peaks in the photon energy density appear around the location where $\varepsilon - \sqrt{p^2_*-2m_a^2\Phi(z)} = 0$, which happens for 
$\zeta= \pm 72.6$ for the parameter values chosen in the plot. At these locations, the detuning distance is larger than the resonance length ($\Delta \zeta = 4.3$). Hence, for this example, the resonance can develop at these locations. The width of the peaks is roughly $\Delta\zeta$, and the height of the peaks growths in the same way as it would  for the homogeneous axion case in the absence of a gravitational potential.


\begin{figure}[t!]
  \centering
  \includegraphics[width=0.6\linewidth]{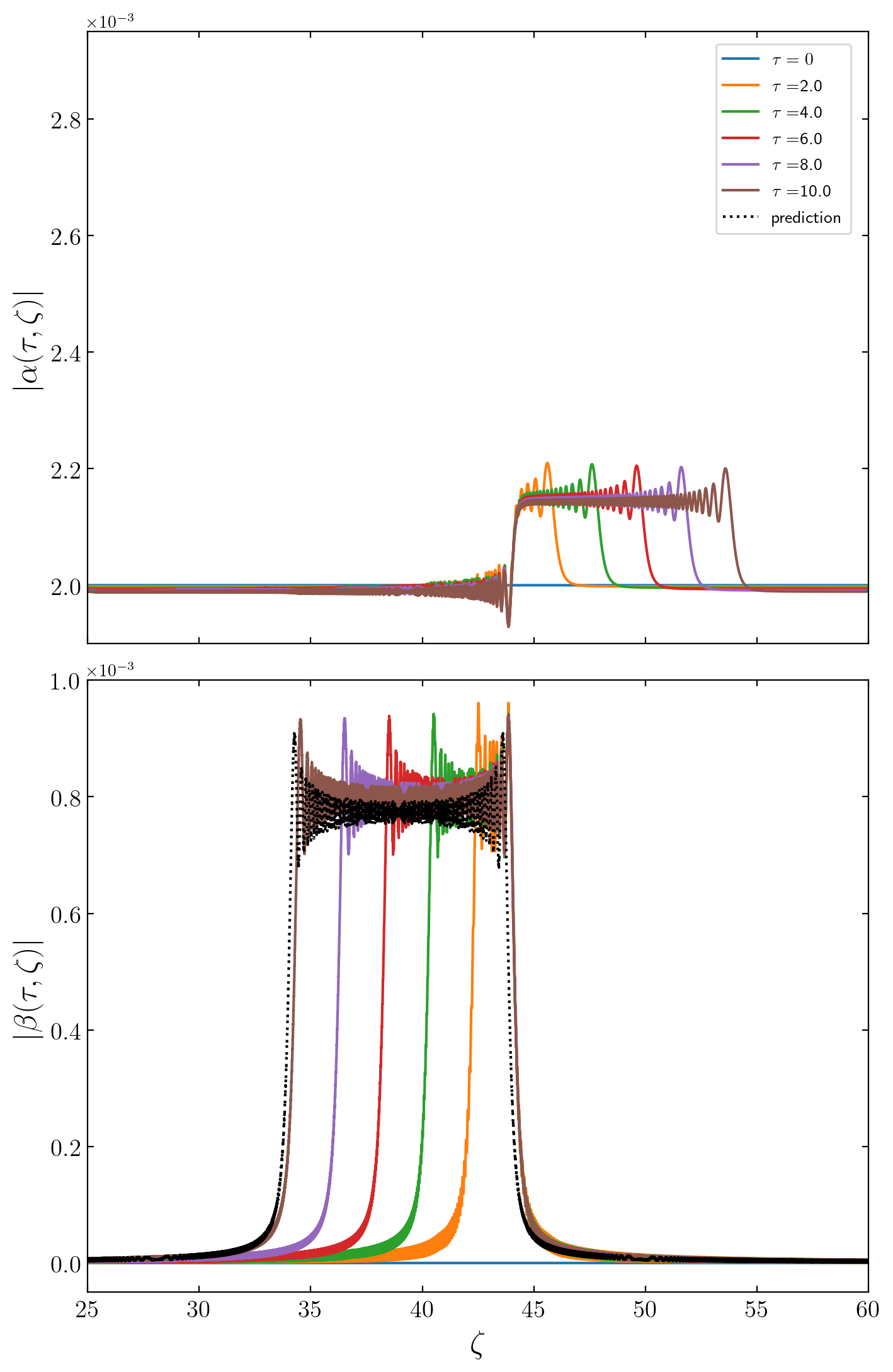}
	\caption{Photon amplitudes at different times with the gravitational potential Eq.~\eqref{gpot} and the background axion field given by Eq.~\eqref{Gsola1}. Parameters are $\tilde{L}=500$, $\tilde{m}_a=35000$, $\tilde{p}_* = 0$, $\alpha_k(0) =\delta_{k0}$, $\beta_k(0)=0$. The parameter $\tilde{\varepsilon}$ is set to 1858.12, such that the peaks in the photon amplitudes initially appear at $\zeta= \pm 43.9$. At this location the detuning distance is $\Delta \zeta = 0.0951$.  The dotted line is the prediction from Eq.~\eqref{sssolbeta13} at $\tau=10$.} 
	\label{fig:steady_grav}
\end{figure}%

In Fig.~\ref{fig:steady_grav}, we show an example where the detuning distance is less than the resonance length: at the location of the resonance 
$\zeta= \pm 43.9$, we have $\Delta \zeta = 0.0951$.
A peak in the photon amplitude appears. As time goes by, the peaks propagate but do not grow in height, creating a steady state similar to Fig.~\ref{fig:Gsteady}. No exponential instability develops.

We can use similar methods to those in Subsection~\ref{sec:coh_clump}, see Eq.~\eqref{ss1eqbeta11}, to describe this phenomenon analytically. Working in the same perturbative framework as there, the first order equation derived from Eq.~(\ref{eqbetaG}) is
\bb
(\partial_\tau-\partial_\zeta)\beta^{(1)}=-{\cal A}\Theta(\tau)\,e^{-i\left(\tilde\varepsilon\tau+S(\zeta)\right)}. \label{ss2eqbeta11}
\ee
Let's choose $\zeta_0$ as the location where the resonance is matched exactly, i.e.\ $\varepsilon=\sqrt{p_*^2-2m_a^2\Phi(\zeta_0)}$. As the resonance is effective only in a small neighborhood  around $\zeta=\zeta_0$, we write
\bb
\Phi(\zeta)\approx\Phi(\zeta_0)+\partial_\zeta\Phi(\zeta_0)(\zeta-\zeta_0). \label{ssPhiexp}
\ee
We can express the derivative of the potential using the detuning distance $\Delta\zeta$ defined in Eq.~\eqref{detune}.
Then $S(\zeta)$ becomes
\bb
S(\zeta)\approx\tilde\varepsilon\zeta-{2\over\Delta\zeta}(\zeta-\zeta_0)^2-\chi \label{sschi1}
\ee
where $\chi=\tilde\varepsilon\zeta_*-2(\zeta_*-\zeta_0)^2/\Delta\zeta$ and $\zeta_*$ corresponds to the reference point at a large distance where the potential vanishes. Transforming $\beta^{(1)}\rightarrow \beta^{(1)}e^{i\chi}e^{-i\varepsilon(\zeta+\tau)}$, Eq. (\ref{ss2eqbeta11}) becomes 
\bb
(\partial_\tau-\partial_\zeta)\beta^{(1)}=-{\cal A}\Theta(\tau)\,e^{{2i\over\Delta\zeta}(\zeta-\zeta_0)^2}. \label{ss2eqbeta12}
\ee
The solution is given by (see Appendix \ref{steady_detail})
\bb
\beta^{(1)}(\tau,\zeta)=-{{\cal A}\over2}\sqrt{\pi i\Delta\zeta\over2}\,
\left[\text{erf}\left(\sqrt{2\over i\Delta\zeta}(\tau+\zeta-\zeta_0)\right)-\text{erf}\left(\sqrt{2\over i\Delta\zeta}(\zeta-\zeta_0)\right)\right] \label{sssolbeta13}
\ee
The steady state is obtained for large times, when 
$\tau\gg \sqrt{\Delta\zeta}$, or, in physical units, $t \gg \sqrt{\Delta z/\sigma}$. In this limit, we get
\bb
\beta_\text{steady}^{(1)}(\zeta)=-{{\cal A}\over2}\sqrt{\pi i\Delta\zeta\over2}\,
\left[1-\text{erf}\left(\sqrt{2\over i\Delta\zeta}(\zeta-\zeta_0)\right)\right] \enspace. \label{sssolbeta14}
\ee
The black dotted line in the lower panel of Figure~\ref{fig:steady_grav} shows the prediction Eq.~\eqref{sssolbeta13}, evaluated at $\tau=10$, against the result of a numerical simulation with $\Delta\zeta = 0.00951$ at the location where the peaks initially appear. We can see that there is very good agreement.

\begin{figure}[t!]
  \centering
  \includegraphics[width=0.6\linewidth]{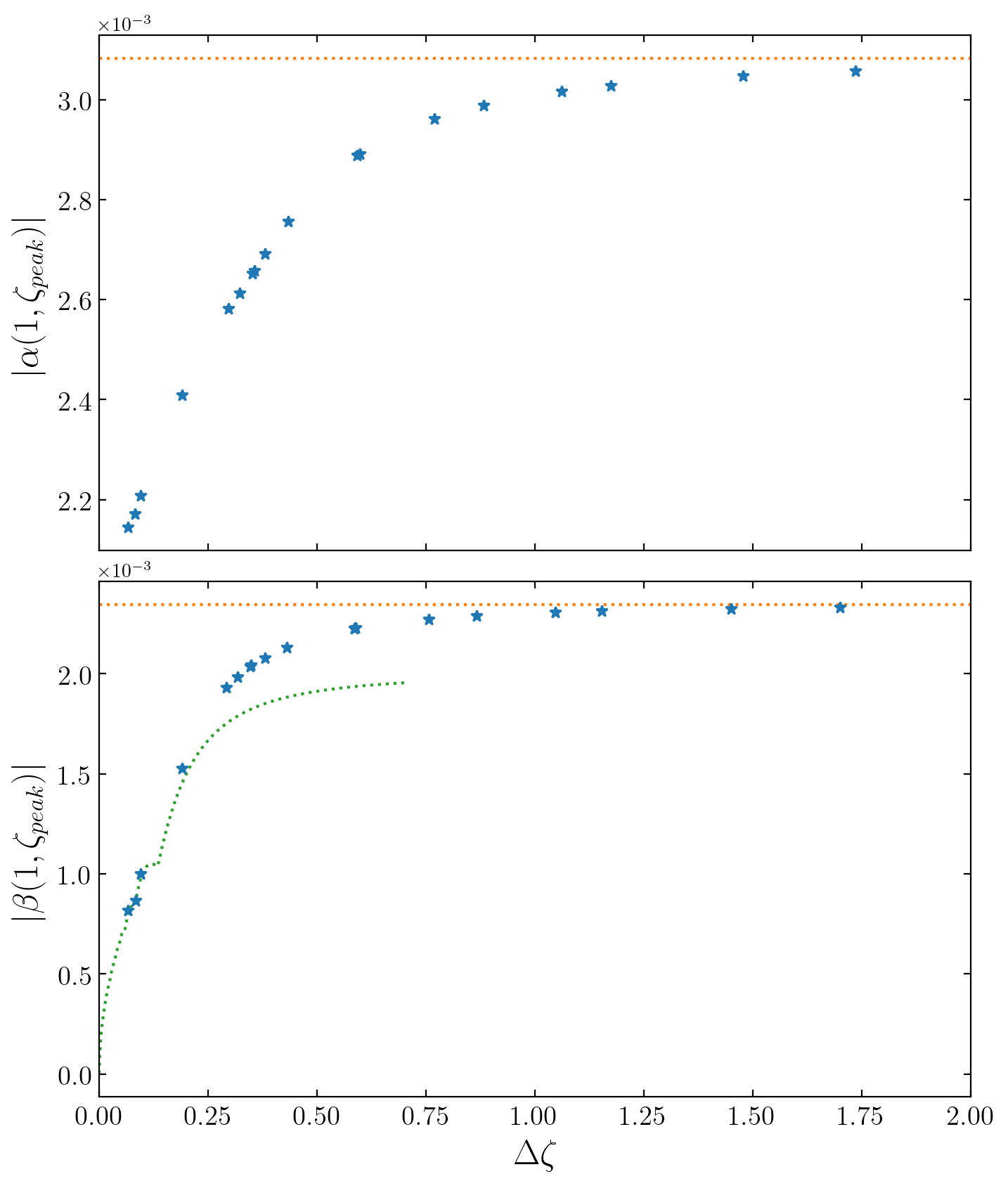}
	\caption{Peak height of the absolute value of the photon amplitudes at time $\tau=1$ as a function of $\Delta \zeta$. 
	Eq.~\eqref{detune} at the location corresponding to the center of the peak. The initial value of $|\alpha|$ is 0.002. 
	 The dotted orange lines represent the value the amplitudes would have in the absence of gravitational potential.
	 The dotted green line is the maximum of Eq.~\eqref{sssolbeta13}, for the corresponding $\Delta\zeta$ and $\tau=1$.} 
	\label{fig:stars}
\end{figure}%

Figure~\ref{fig:stars} shows the peak height at $\tau =1$ as a function of  $\Delta\zeta$ at the location where the peak appears. The horizontal orange lines represent the value the amplitudes would have at $\tau=1$ in the absence of a gravitational potential. We see that for $\Delta\zeta \gtrsim 1$, the peak height reaches the orange line. We are thus in the regime of exponential growth. If we were to run the simulation for a longer time, the peak heights would keep growing exponentially as the waves propagate, similar to what observed in Fig.~\ref{fig:Gexp}. For  $\Delta\zeta \lesssim 1$, we enter the steady state regime. The dotted green line represents the maximum value of $\beta^{(1)}(1,\zeta)$ of Eq.~\eqref{sssolbeta13}. For $\Delta\zeta \lesssim 0.25$, there is excellent agreement between prediction and simulations. This implies that the photon amplitudes represented by the first four stars have already reached their maximum value at $\tau =1$, and thus the peak height would not grow further if we were to run the simulations longer. 
From Eq.~\eqref{sssolbeta14}, for $\zeta\rightarrow -\infty$,  we obtain that, in the limit $\Delta\zeta\rightarrow 0$, the height of the wave front goes to zero as $\sqrt{\Delta\zeta}$.

\section{Discussion and conclusions}\label{sec:disc}

Photons traveling through an axion background can experience parametric resonance, which can potentially lead to an exponential growth of the photon field. Microscopically this effect can be interpreted as stimulated axion decay into two photons. In this paper, we have studied quantitatively the impact of a non-trivial momentum distribution of the axion background field and the effect of a gravitational potential, which can lead to a detuning of the resonance due to the gravitational redshift.

The characteristic scale of the resonance is set by the growth factor $\sigma$ given in Eq.~\eqref{eq:mu-bar}. It is proportional to the axion--photon coupling $\gag$ and the square-root of the local axion energy density, and it
can be interpreted as an inverse resonance length.
 The main results of our work are the following two conditions which must  be fulfilled in order for an exponential instability to develop:
\begin{enumerate}
    \item The momentum spread $\delta_p$ of the axion field has to be smaller than the inverse resonance length: $\sigma > \delta_p$. 
    \item The detuning distance due to the gravitational redshift $\Delta z$ has to be longer than the resonance length: $\sigma \Delta z > 1$.
\end{enumerate}
The first condition requires extremely cold axions. In the case of a coherent clump of axions, condition~1 implies that the size of the clump must be much larger than the resonance length. In the case of a random field, condition~1 implies that the coherence length of the axion must be longer than the resonance length. For typical values of axion dark matter densities in our galaxy, the resonance length is of order $1 \, \rm pc\sim 10^{23} \, eV^{-1}$, see Eq.~\eqref{sigma_value}. In the common picture of virialized axions forming the galaxy, one expects the coherence length to be set by the de Broglie wavelength of the axions, which is of order $1/(v m_a)$ with typical velocities $v\sim 10^{-3}$. Hence, for axions with masses such that the resonance happens in the observable frequency range, $m_a \gtrsim 10^{-8}$~eV, condition~1 is violated by many orders of magnitude, and no resonance can develop for a virialized axion dark matter halo.  Note that for higher virial velocities (e.g., for axions bound in smaller, sub-galactic substructures), the condition is even more strongly violated than for our fiducial example of $v\sim 10^{-3}$.

The second condition requires extremely weak or flat gravitational potentials. We have shown that under the assumption of axion dark matter, for typical values, condition~2 is violated strongly in our galaxy. Therefore, no resonance is possible in the background of galactic dark matter axions, even if we assume the axions to form an extremely cold condensate.

Let us now provide some rough order of magnitude estimates of the two conditions applied to a clump of axions with total mass $M$ and characteristic size $R$. We estimate $\delta_p \sim 1/R$, $\rho_a \sim M/R^3$, and using the definition of  
$\sigma$ in Eq.~\eqref{eq:mu-bar}, condition~1 becomes
\bq
R < \frac{\gag^2}{8} M 
\sim 10^5\,{\rm m} \,
\left( \frac{\gag}{10^{-11}~\mathrm{GeV}^{-1}} \right)^2
\left( \frac{M}{10^{-13}~M_\odot}\right)\enspace.
\eq
Condition~2 is given by $\Delta\zeta > 1$, with $\Delta\zeta$ defined in Eq.~\eqref{detune}. We can estimate the potential and its derivative by dimensional analysis based on Poisson's equation: $\Phi \sim -4\pi G_N \rho_a R^2$, $\partial_z \Phi \sim \Phi/R$. Dropping numerical factors, the condition becomes
\bq
R < \left(\frac{\gag^4 M}{m_a^2 G_N}\right)^{1/3} 
\sim 10^6\,{\rm m} \,
\left( \frac{\gag}{10^{-11}~\mathrm{GeV}^{-1}} \right)^{4/3}
\left( \frac{M}{10^{-13}~M_\odot}\right)^{1/3}
\left( \frac{10^{-5}~\mathrm{eV}}{m_a}\right)^{2/3}
\enspace.
\eq
Note that for $M\sim 10^{-13}\, M_\odot$, which is a typical mass of axion stars, the two conditions are roughly of similar strength, with some dependence on the axion mass. For smaller clump masses, the momentum condition dominates whereas for larger masses the redshift requirement is more important. We leave a detailed investigation of the parametric resonance for axion stars for future work \cite{future}.

Let us stress that the conditions~1 and 2 above are necessary conditions for the instability to develop but certainly not sufficient. For instance, an effective photon mass induced by medium effects can additionally prevent the axion decay, see e.g., \cite{Alonso-Alvarez:2019ssa} for a recent discussion in the cosmological context. 

Finally, it is worth mentioning that the echo method for axion dark matter detection \cite{Arza:2019nta} is not hindered by our analysis. The echo does not rely on the exponential regime of the parametric resonance, but it is formed in the perturbative regime. 

\bigskip

{\it Acknowledgement.} We would like to thank Andrea Caputo, Saptarshi Chaudhuri, Andreas Pargner, Nicholas Rapidis, Bj\"orn-Malte Sch\"afer, Pierre Sikivie and Karl van Bibber for useful discussions.  This project has been supported by the European Unions Horizon 2020 research and innovation program under the Marie Sklodowska-Curie grant agreement No 674896 (Elusives). T.S.\ and E.T.\ acknowledge support by the Munich Institute for Astro- and Particle Physics (MIAPP), which is funded by the Deutsche Forschungsgemeinschaft (DFG, German Research Foundation) under Germany's Excellence Strategy EXC-2094-390783311. A. A.\ thanks the University of Santiago of Chile for their hospitality.

\appendix
\section{Eikonal approximation for the axion in a weak gravitational potential}\label{eikonal}

We look for a solution to Eq.~\eqref{eom_axion} in the form 
\bb
a(t,\x\,) = (\text{const})\times e^{i S(t,\x\,)} \enspace,
\label{eq:eikonal-ansatz}
\ee
where $S$ is a scalar under coordinate transformations. We use the ansatz above  at zeroth order in the eikonal approximation: $\partial^2 S \ll (\partial S)^2$, for derivatives in any space-time direction.
In this approximation, the equation for $S$ is 
\bb
(1-2\Phi) (\partial_0 S)^2  - (1+2\Phi) \delta^{ij} (\partial_i S)(\partial_j S) - m_a^2 = 0 \enspace.\label{eqS}
\ee
We define the covariant axion four-momentum as 
\bb
p_\mu = \partial_\mu S  \enspace.
\label{eq:defpmu}
\ee
Then Eq.~\eqref{eqS} becomes just the dispersion relation  $g^{\mu\nu}p_\mu p_\nu = m_a^2$.
Applying a covariant derivative $\nabla_\alpha$ to the dispersion relation and using the fact that ${\nabla_\alpha p_\nu = \nabla_\nu p_\alpha}$ (which follows from \eqref{eq:defpmu} and the symmetry of the Christoffel symbols in the lower indices), one can show that $p_\mu$ solves the geodesic equation $p^\mu\nabla_\mu p^\nu =0$.

For the metric~\eqref{Gmetric}, the non-null Christoffel symbols to linear order in $\Phi$ are
\bb
\Gamma^0_{0i} = \partial_i \Phi \,,
\qquad
\Gamma^i_{00} = \delta^{ij} \partial_j \Phi \,,
\qquad
\Gamma^i_{jk} = -\delta^i_j \partial_k \Phi  -\delta^i_k \partial_j \Phi + \delta_{jk}\delta^{il} \partial_l \Phi \enspace.
\ee
The proper time is $d\tau^2 = ds^2$.
The geodesic equations for the  four-velocity $u^\mu = dx^\mu/d\tau$ are
\bq
\frac{du^0}{d\tau} &=& -2\partial_i\Phi\ u^0 u^i \label{geo0}\\
\frac{du^i}{d\tau} &=& -\delta^{ij}\partial_j\Phi\ u^0 u^0 +2 \delta^i_j \partial_k  \Phi \ u^j u^k 
- \delta_{jk}\delta^{il} \partial_l \Phi \ u^j u^k   \enspace.\label{geoi}
\eq
We express $\tau$ as a function of $x^0=t$. 
We can then rewrite Eq.~\eqref{geo0} as 
\bb
\frac{du^0}{dt} = -2\partial_i\Phi\ u^i \enspace.
\ee
Using
\bb
u^i = \frac{dx^i}{d\tau} = u^0\frac{dx^i}{dt}   \enspace,
\ee
we get
\bb
\frac{du^0}{u^0} = -2\ d\Phi \enspace.
\ee
To linear order in $\Phi$, the solution to the equation above is
\bb
u^0 = (1-2\Delta\Phi)u^0_* \enspace,
\ee
where we have chosen a reference point $x^i_*$ where $\Phi = \Phi_*$ and $u^0 = u^0_*$ and $\Delta\Phi = \Phi - \Phi_*$.

To solve Eq.~\eqref{geoi}, we write $u^i = \hat{n}^i u$, where $u$ is defined as $u=\sqrt{-\eta_{ij}u^iu^j}$, implying $\eta_{ij}\hat{n}^i\hat{n}^j=-1$.
Notice that, since the gravitational potential is a perturbation to Minkowski space, we take $\hat{n}^i$ to lie along the background trajectory.
We obtain an equation for $u$ by multiplying~\eqref{geoi} by $\delta_{im}\hat{n}^m$. We get
\bb
\frac{du}{d\tau} = -\hat{n}^i\partial_i\Phi \qquad\Rightarrow\qquad  
du = -\frac{1}{u^0}\hat{n}^i\partial_i\Phi dt = -\frac{1}{u}\ d\Phi 
 \enspace.
\ee
The solution is
\bb
u^2 = u^2_* - 2\Delta\Phi \qquad\Rightarrow\qquad  u = \sqrt{u^2_* - 2\Delta\Phi }\enspace.
\ee
In terms of $p$, the result is
\bq
p^0 &=& (1-2\Delta\Phi)p^0_* \label{eq:p0_contra} \\
p^i &=& \hat{n}^i\sqrt{\left(p_*^j\right)^2 - 2m_a^2\Delta\Phi }\enspace.
\eq
We lower the indices using the metric:
\bq
p_0 &=& p_{0*} \label{eq:p0_cov} \\
p_i &=& \hat{n}_i (1-2\Delta\Phi)
\sqrt{(p_{j*})^2 - 2m_a^2\Delta\Phi } \enspace,
\eq
where $\hat{n}_i = \hat{n}^i$, 
$(p_{j*})^2 \equiv \delta^{ij} p_{i*}p_{j*}$ and
$(p_{*}^j)^2 \equiv \delta_{ij} p_{*}^i p_{*}^j$.
Let us chose the $z$ direction parallel to $\hat{n}^i$. Then the axion phase is obtained from Eq.~\eqref{eq:defpmu}:
\bb
S(t, z) = \pm p_0 (t-t_*) \pm \int^{z}_{z_*} dz'\, (1-2\Delta\Phi)\sqrt{(p_{j*})^2 - 2m_a^2\Delta\Phi } \enspace.\label{Sa}
\ee
We now chose the reference point $*$ at infinity, with $\Phi_*=0$. Then, we have $\Delta\Phi = \Phi$ and
$(p_{j*})^2 = (p_{*}^j)^2 \equiv  p_{*}^2$. 
Assuming further that $\Phi$ depends only on $z$,
we can perform the coordinate transformation
$dz\rightarrow dz/(1-2\Phi)$ and recover the axion phase used in Eqs.~\eqref{Gsola1} and \eqref{GS1}.

Notice that the geodesic equation for photons in the eikonal approximation leads to an identical redshift effect for the zero-component of the photon 4-momentum as in Eq.~\eqref{eq:p0_contra} or \eqref{eq:p0_cov}. Hence, the \emph{frequency} of the photon and the axion fields are affected in the same way by the redshift. In this respect, we do not agree with the arguments given in Ref.~\cite{Wang:2020zur}. The crucial effect appears due to the \emph{spatial} component of the 4-momentum, which behaves differently under the influence of gravity for relativistic and non-relativistic particles, leading to the detuning of the resonance.

\section{WKB approximation for gravitationally bound axions}\label{wkb}

We briefly sketch the derivation of stationary solutions of Eq.~\eqref{geqa1} involving standard quantum mechanics methods. 
Since we are interested in non-relativistic axions, we can apply the standard procedure to transform Eq.~\eqref{geqa1} into a Schr\"odinger equation (see e.g.\ appendix of \cite{Davidson:2016uok}), by
writing the axion field as 
\bb
a = \frac{1}{2i} \left( \alpha e^{-im_at} - \text{c.c.} \right)
\ee
with  $\partial_t\alpha \ll m_a$ and $\partial_t^2\alpha \ll m_a\partial_t\alpha$. To leading order, we get 
\bb
i\partial_t\alpha = -\frac{1}{2m_a}\partial_z^2\alpha + m_a\Phi\alpha \enspace. 
\ee
Now, we write $\alpha(t, z) = \psi(z) e^{-iEt}$ to obtain the time independent Schr\"odinger equation for the space part:
\bb
\frac{1}{2m_a} \partial_z^2\psi = (m_a\Phi - E)\psi \enspace.
\ee
Note that the slowly varying assumption for $\alpha$ implies $|E| \ll m_a$. Now we can use in addition that the gravitational potential is slowly varying in the sense of Eq.~\eqref{eq:approx1}, in which case an approximate solution for $\psi$ is given by the WKB solution, see e.g.~\cite{sakurai},
\bb
\psi(z) \approx u(z)
\,\exp\left[\pm i \int^z dz'\sqrt{2m_a[E-m_a\Phi(z')]} \right] 
\enspace, \label{eq:wkb-app}
\ee
where $u(z)$ is a slowly varying function. We ignore the $z$ dependence of $u$ and treat it as a constant. This amounts to working at zeroth order in the WKB approximation, consistent with the eikonal approximation in Eq.~\eqref{eq:eikonal-ansatz}. In other words, we include the effect of the potential only in the phase of the axion field, but neglect it for the amplitude. 
The axion field adopted in  Eqs.~\eqref{Gsola1} and \eqref{GS1} is obtained from Eq.~\eqref{eq:wkb-app} with $p_*^2 = 2m_aE$ and by setting $m_a + E \approx m_a$ in the time dependent exponent.

Let us comment briefly on the motivation for the WKB ansatz and the values of the binding energy $E$. In general, $E$ is quantized, and there is a minimal ground state energy $E_0$. Here, we are interested in axion configurations coherent over distances much larger than the resonance length, which, for typical values, in our galaxy is of order 1~pc, see Eq.~\eqref{sigma_value}. Furthermore, we focus on photons in the observable range, implying axion masses corresponding to Compton wavelengths smaller than $\mathcal{O}(10)$~m, i.e., $m_a \gtrsim 10^{-8}$~eV. With a potential $\Phi$ motivated by the gravitational potential of the galaxy, those conditions can only be satisfied for very loosely bound states with $E \gg E_0 \sim m_a \Phi(z=0)$. Furthermore, demanding that the dark matter halo extends much further out than our location in the galaxy requires also that the classical turning point at $E=m_a\Phi$ is at much larger radii. This implies that $E\gg m_a \Phi(z=z_\oplus)$ (remember, both $E$ and $\Phi$ are negative). This justifies the approximation $|p_*|/m_a \ll \sqrt{-2\Phi}$ adopted for our numerical estimates as well as the WKB approximation in Eq.~\eqref{eq:approx1}.

\section{Details of the numerical simulations}\label{num_detail}

The discrete form of Eqs.~(\ref{eqalpha},~\ref{eqbeta}) is
\bq
\partial_\tau \alpha_r
&=&  
- i\left(\frac{\tilde{\varepsilon}}{2} + \Delta  r \right) \alpha_r
-\frac{1}{N} \sum_{s=-N/2}^{N/2-1} \beta_s\ F^*(-r+s)\label{eq_matrix_alpha}
\\
\partial_\tau \beta_r 
&=& 
 i\left(\frac{\tilde{\varepsilon}}{2} + \Delta  r \right) \beta_r 
-\frac{1}{N} \sum_{s=-N/2}^{N/2-1}\alpha_s\ F(r-s)\enspace, \label{eq_matrix_beta}
\eq
where
\bb
F(u) =
\sum_n f_n \  \e^{-i \frac{2\pi}{N} u n} \enspace,
\ee
and $r$ and $s$  are discrete indices.
We write Eqs.~(\ref{eq_matrix_alpha},~\ref{eq_matrix_beta}) in matrix form
\bq
\partial_\tau \alpha
&=&  - D \alpha + L^\dagger \beta
\\
\partial_\tau \beta 
&=& D \beta + L \alpha \enspace,
\eq
where 
\bq
\alpha &=& (\alpha_{-N/2}, \alpha_{-(N+1)/2}, ..., \alpha_{(N-1)/2}, \alpha_{N/2})^T\\
\beta &=& (\beta_{-N/2}, \beta_{-(N+1)/2}, ..., \beta_{(N-1)/2}, \beta_{N/2})^T\enspace,
\eq
and $D$, $L$ are $N\times N$ matrices following from Eqs.~(\ref{eq_matrix_alpha},~\ref{eq_matrix_beta}).
The matrix $L$ has $(N/2-1)^2$ elements of the form  $F(u )$ with $|u|> N/2$. We set those elements to zero, as they correspond to frequencies higher the Nyquist frequency of our simulation. The matrix $L$ is then
\bq
\tiny
L= 
-\frac{1}{N}
\left(
\begin{array}{ccccccccccc}
F(0) & F(-1)&\hdots & F(-\frac{N}{2}+1) &F(-\frac{N}{2}) &0&\hdots & 0 &0 \\
F(1) & F(0)&\hdots & F(-\frac{N}{2}+2) &F(-\frac{N}{2} + 1 ) &F(-\frac{N}{2})&\hdots & 0&0 \\ \\ 
\vdots & \vdots &\hdots & \vdots & \vdots  & \vdots &\hdots & \vdots & \vdots \\ \\ 
F(\frac{N}{2}-1) & F(\frac{N}{2}-2)&\hdots & F(0) &F(-1) &F(-2)&\hdots & F(-\frac{N}{2}+1) &F(-\frac{N}{2})\\ 
F(\frac{N}{2}) & F(-\frac{N}{2}-1)&\hdots & F(1) &F(0) &F(-1)&\hdots & F(-\frac{N}{2}+2) &F(-\frac{N}{2}+1)\\ 
0 & F(\frac{N}{2})&\hdots & F(2) &F(1) &F(0)&\hdots & F(-\frac{N}{2}+3) &F(-\frac{N}{2}+2)\\ \\
\vdots & \vdots &\hdots & \vdots & \vdots  & \vdots &\hdots & \vdots & \vdots \\ \\ 
0 & 0&\hdots & F(\frac{N}{2}-1) &F(\frac{N}{2} - 2 ) &F(\frac{N}{2}-3)&\hdots & F(0) &F(-1)\\
0 & 0&\hdots & F(\frac{N}{2}) &F(\frac{N}{2} - 1 ) &F(\frac{N}{2}-2)&\hdots & F(1) &F(0) \\
\end{array} \ \ \right) \enspace.
\eq
This matrix is of the Toeplitz type. 
We embed $L$ in a circulant matrix $C$
\bq
C =
\left(
\begin{array}{cc}
L & Aux\\
Aux & L
\end{array} \ \ \right) \enspace, 
\eq
where $Aux$ is an auxiliary $N\times N$ matrix whose first column is
\bq
\left(
0 ,\ 0,\ \hdots ,\ 0,\ F(-N/2),\ F(-N/2 +1),\ \hdots ,\ F(-2),\ F(-1) \right)^T  \enspace.
\eq
Any matrix product with a vector $C.v$ can be computed as $\mathcal{F}^{-1}\Lambda \mathcal{F}v $, where $\mathcal{F}$ is the Fast Fourier Transform (FFT) operation and $\Lambda$ is a diagonal matrix whose elements are the FFT of the first column of $C$. Performing the matrix product this way considerably speeds up computations if $N$ is a power of 2.

\section{Calculation of the steady state transition} \label{steady_detail}

In this Section, we show the detailed solutions of Eqs. (\ref{ss1eqbeta11}) and (\ref{ss2eqbeta12}). Both equations can be written in the form
\bb
(\partial_\tau-\partial_\zeta)\beta^{(1)}=-{\cal A}\Theta(\tau)e^{-\eta(\zeta-\zeta_0)^2}\enspace. \label{sdeqbeta1}
\ee
The general solution for (\ref{sdeqbeta1}) is given by
\bb
\beta^{(1)}(\tau,\zeta)=-{\cal A}\int_{-\infty}^{\infty}d\zeta'\int_{-\infty}^{\infty}d\tau'\,G(\tau-\tau',\zeta-\zeta')\,\Theta(\tau')\,e^{-\eta(\zeta'-\zeta_0)^2}\enspace, \label{sdsolbeta1}
\ee
where $G(\tau,\zeta)$ is the retarded Green's function associated with the operator $\partial_\tau-\partial_\zeta$.  $G(\tau,\zeta)$ satisfies
\bb
(\partial_\tau-\partial_\zeta)G(\tau,\zeta)=\delta(\tau)\delta(\zeta)\enspace. \label{sdeqG1}
\ee
To solve Eq. (\ref{sdeqG1}), we expand $G(\tau,\zeta)$ in Fourier space as 
\bb
G(\tau,\zeta)={1\over(2\pi)^2}\int_{-\infty}^\infty\int_{-\infty}^\infty\, dk\,d\omega\, e^{i(k\zeta-\omega\tau)}\hat G(\omega,k)\enspace. \label{sdsolG1}
\ee
Plugging (\ref{sdsolG1}) into (\ref{sdeqG1}), we get
\bb
\hat G(\omega,k)={i\over\omega+k}\enspace, \label{sdsolGF1}
\ee
and therefore
\bb
G(\tau,\zeta)={i\over(2\pi)^2}\int_{-\infty}^\infty dk\,e^{ik\zeta}\int_{-\infty}^\infty\, d\omega{e^{-i\omega\tau}\over\omega+k}\enspace. \label{sdsolG2}
\ee
To find the retarded solution, we want $G(\tau,\zeta)$ to vanish for $\tau<0$. To do that we shift $\omega\rightarrow\omega+i\epsilon$ in the integration of (\ref{sdsolG2}). Using the Residue theorem, we find $G(\tau,\zeta)=0$ for $\tau<0$ and
\bq
G(\tau,\zeta) &=& {i\over(2\pi)^2}\int_{-\infty}^\infty dk\,e^{ik\zeta}(-2\pi i)e^{ik\tau} \nonumber
\\
&=& {1\over2\pi}\int_{-\infty}^\infty dk\, e^{ik(\zeta+\tau)} \nonumber
\\
&=& \delta(\tau+\zeta) \enspace,\nonumber
\eq
for $\tau>0$. So we have
\bb
G(\tau,\zeta)=\Theta(\tau)\delta(\tau+\zeta)\enspace. \label{sdsolG3}
\ee
Now we proceed to integrate (\ref{sdsolbeta1}). For $\tau>0$ we have
\bq
\beta^{(1)}(\tau,\zeta) &=& -{\cal A}\int_{-\infty}^{\infty}d\zeta'\,e^{-\eta(\zeta'-\zeta_0)^2}\int_{-\infty}^{\infty}d\tau'\,\Theta(\tau')\Theta(\tau-\tau')\,\delta(\tau-\tau'+\zeta-\zeta') \nonumber
\\
&=& -{\cal A}\int_{-\infty}^{\infty}d\zeta'\,e^{-\eta(\zeta'-\zeta_0)^2}\Theta(\tau+\zeta-\zeta')\Theta(\zeta'-\zeta) \nonumber
\\
&=& -{\cal A}\int_{\zeta}^{\zeta+\tau}d\zeta'\,e^{-\eta(\zeta'-\zeta_0)^2} \nonumber
\\ 
&=& -{{\cal A}\over2}\sqrt{\pi\over\eta}\left(\text{erf}(\sqrt{\eta}(\tau+\zeta-\zeta_0))-\text{erf}(\sqrt{\eta}(\zeta-\zeta_0))\right)\enspace.   \label{sdsolbeta2}
\eq
The steady state solution is found for $\tau\rightarrow\infty$. For large $x$, $\text{erf}(\sqrt{\eta}x)$ goes to $1$ as long as $\text{Re}(\eta)\leq0$. Since that is the case for Eqs. (\ref{ss1eqbeta11}) and (\ref{ss2eqbeta12}), we find
\bb
\beta_\text{steady}^{(1)}(\zeta)=-{{\cal A}\over2}\sqrt{\pi\over\eta}\left(1-\text{erf}(\sqrt{\eta}(\zeta-\zeta_0))\right)\enspace.   \label{sdsolbeta3}
\ee

\bibliographystyle{JHEP_improved}
\bibliography{./refs}

\end{document}